\documentclass[conference]{IEEEtran}
\IEEEoverridecommandlockouts
\def\BibTeX{{\rm B\kern-.05em{\sc i\kern-.025em b}\kern-.08em
    T\kern-.1667em\lower.7ex\hbox{E}\kern-.125emX}}

\usepackage{amsmath}

\usepackage{fixltx2e}
\usepackage{hyperref}
\usepackage{textcomp}
\usepackage{listings}
\usepackage{array}
\usepackage[utf8]{inputenc}
\usepackage{xcolor}
\usepackage{multirow}
\usepackage[obeyspaces]{xurl}
\usepackage{graphicx}
\usepackage{xspace}
\usepackage{caption}
\usepackage{amssymb}
\usepackage{algorithm}
\usepackage{algorithmic}
\usepackage{hhline}
\graphicspath{ {figures/} }

\definecolor{codegreen}{rgb}{0,0.6,0}
\definecolor{codegray}{rgb}{0.5,0.5,0.5}
\definecolor{codepurple}{rgb}{0.58,0,0.82}
\definecolor{backcolour}{rgb}{0.95,0.95,0.92}

\lstdefinestyle{mystyle}{
    backgroundcolor=\color{backcolour},
    commentstyle=\color{codegreen},
    keywordstyle=\color{magenta},
    numberstyle=\tiny\color{codegray},
    stringstyle=\color{codepurple},
    basicstyle=\ttfamily\footnotesize,
    breakatwhitespace=false,
    breaklines=true,
    captionpos=b,
    keepspaces=true,
    numbers=left,
    numbersep=5pt,
    showspaces=false,
    showstringspaces=false,
    showtabs=false,
    tabsize=2
}

\newcommand{\syspp}[0]{P{$^2$}IM\xspace}
\newcommand{\sysname}{DICE\xspace} %
\newcommand{\sysnamelong}{DMA Input Channel Emulation\xspace} %

\newcommand{\mipspic}{PIC32 emulator\xspace}

\newcommand{\point}[1]{\vspace{0.1in}\par\noindent{\bf #1:}}

\newcommand{\eg}{\textit{e.g.,}\xspace}
\newcommand{\ie}{\textit{i.e.,}\xspace}
\newcommand*{\rom}[1]{\textit{\expandafter\romannumeral #1}}

\newcommand{\din}{DMA input channel\xspace}
\newcommand{\dins}{DMA input channels\xspace}
\newcommand{\dout}{DMA output channel\xspace}
\newcommand{\douts}{DMA output channels\xspace}
\newcommand{\dctr}{DMA controller\xspace}
\newcommand{\dctrs}{DMA controllers\xspace}
\newcommand{\dstr}{DMA stream\xspace} %
\newcommand{\dstrs}{DMA streams\xspace} %
\newcommand{\dtran}{DMA transfer\xspace} 
\newcommand{\dtrans}{DMA transfers\xspace} 
\newcommand{\ddesr}{transfer descriptor\xspace} %
\newcommand{\ddesrs}{transfer descriptors\xspace} %
\newcommand{\dptr}{transfer pointer\xspace} %
\newcommand{\dptrs}{transfer pointers\xspace} %

\begin{document}

\title{\sysname: Automatic Emulation of DMA Input Channels for Dynamic Firmware Analysis}

\author{\IEEEauthorblockN{Alejandro Mera, Bo Feng, Long Lu, Engin Kirda, William Robertson}
\IEEEauthorblockA{\textit{Khoury College of Computer Sciences} \\
\textit{Northeastern University}\\
 Boston, USA \\
 \{mera.a, feng.bo, l.lu, e.kirda\}@northeastern.edu }}

\maketitle
\thispagestyle{plain}
\pagestyle{plain}

\begin{abstract}

Microcontroller-based embedded devices are at the core of Internet-of-Things
(IoT) and Cyber-Physical Systems (CPS). The security of these devices is of
paramount importance. Among the approaches to securing embedded devices, dynamic
firmware analysis (e.g., vulnerability detection) gained great attention lately,
thanks to its offline nature and low false-positive rates. However, regardless
of the analysis and emulation techniques used, existing dynamic firmware
analyzers share a major limitation, namely the inability to handle firmware
using DMA (Direct Memory Access). It severely limits the types of devices
supported and firmware code coverage. 

We present \sysname, a drop-in solution for firmware analyzers to emulate \dins
and generate or manipulate DMA inputs (from peripherals to firmware). \sysname
is designed to be hardware-independent (i.e., no actual peripherals or DMA
controllers needed) and compatible with common MCU firmware (i.e., no
firmware-specific DMA usages assumed) and embedded architectures. The high-level idea behind \sysname is
the identification and emulation of the abstract \dins, rather than the highly
diverse peripherals and controllers. 
\sysname identifies \dins as the firmware writes the source and destination DMA \dptrs into the \dctr. 
Then \sysname manipulates the input transferred through DMA on behalf of the firmware analyzer.
\sysname does not require firmware source
code or additional features from firmware analyzers.

We integrated \sysname to the recently proposed firmware analyzer \syspp (for ARM Cortex-M architecture) and
a \mipspic (for MIPS M4K/M-Class architecture). 
We evaluated it on 83 benchmarks and sample firmware, representing 9 different \dctrs from
5 different vendors. \sysname detected 33 out of 37 \dins, with 0 false positives. It correctly supplied
DMA inputs to 21 out of 22 DMA buffers that firmware actually use, 
which previous firmware analyzers cannot achieve due to the lack of DMA emulation. \sysname's 
overhead is fairly low, it adds 3.4\% on average to \syspp execution time. We also 
fuzz-tested 7 real-world firmware using \sysname and compared the results with
the original \syspp. \sysname uncovered tremendously more execution 
paths (as much as 79X) and found 5 unique previously-unknown bugs that are 
unreachable without DMA emulation. All our source code and dataset are publicly 
available.

\end{abstract}
\section{Introduction}

Modern embedded devices, equipped with increasingly powerful MCUs 
(microcontrollers) and rich network connectivity, are used as the building
blocks in Internet-of-Things (IoT) and Cyber-Physical Systems (CPS). It
is expected that 5.8 billion Enterprise and Automotive connected devices 
(IoT and CPS) will be in use in 2020 \cite{Gartner2020}.
Therefore, the (in)security of embedded devices has profound implications on 
millions of devices, in terms of both data privacy and physical safety. 
Security vulnerabilities in firmware may allow attackers to control affected devices deployed in
smart homes, connected vehicles, intelligent factories, power grids, etc.,
and in turn, steal critical data or manipulate device behavior. Such attacks
have been on a rise and launched on cars \cite{AutomotiveCheckoway,RemoteCarHackMiller}, 
Wi-Fi SoC \cite{WifiSelianiMarvellAvastar, WifiPzero},
manufacturing controllers \cite{Stuxnet}, and more.

To improve embedded device security, researchers explored various approaches,
including runtime attack mitigation \cite{HardinSPHSK18}, remote attestation \cite{sun2018oat,SWATT},
and firmware analysis \cite{p2im,avatar,firmadyne,inception}. Compared with other approaches, dynamic
firmware analysis has low false positives, requires no hardware or software
modification, and incurs zero overhead on production devices. Therefore, it is
generally considered more practical. 

A major challenge facing dynamic firmware analysis is the inability to fully run
and test firmware at scale, due to hardware dependence on diverse peripherals.
Many previous works either rely on real hardware components during analysis or
port firmware to a conventional computer platform (e.g., x86), for which full
emulators exist. \syspp~\cite{p2im} is a recent work that for the first time enables dynamic
firmware analysis without requiring actual hardware, source code, or porting firmware 
to a non-native platform. \syspp removes hardware dependence by identifying
processor-peripheral interfaces and supplying viable input data through such
interfaces on behalf of peripherals. As a result, firmware can boot, run, and be
tested in a generic emulator without using peripheral hardware or emulation.

Despite the tremendous progress made by previous works, one fundamental problem
remains open: existing dynamic analyzers cannot support firmware taking input
from peripherals via DMA (Direct Memory Access). When firmware reads from a DMA
buffer in memory, which is supposed to contain input written directly by a
peripheral, existing analyzers or emulators would fail to recognize it as a DMA
read, and instead, treat it as a regular memory read (i.e., returning zero or
invalid value to firmware). As a result, firmware cannot obtain any DMA inputs,
which causes the execution to idle, a large portion of the firmware code to be
unreachable/untested, or even the analysis session to crash.   

The root cause of the problem lies in the very nature of DMA, which allows
peripherals to access memory directly (with the help of a DMA controller, which
is also a peripheral). A firmware analyzer, if not fully emulating all
peripherals or entirely aware of their DMA activities, is unable to determine
when and where in memory DMA-based I/O may occur. Therefore, the analyzer cannot
tell, when firmware reads from a buffer in memory, whether the read operation is
a DMA-based input event or just a regular load of data from memory.

Due to this open problem, existing analyzers either treat DMA to be out of scope
(i.e., not supporting firmware that uses DMA), or use very simple heuristics to
statically infer locations of DMA buffers in memory, which can be highly
inaccurate and incomplete due to the dynamic nature of DMA. 

Supporting DMA input is critical and necessary for dynamic analysis of embedded
firmware, for the following reasons. First, to perform comprehensive dynamic
tests of firmware, all input channels used by firmware need to be covered.
Embedded devices take inputs from a wide range of peripherals through several 
channels, including MMIO (memory-mapped I/O) and DMA. Analysis of DMA-enabled
firmware cannot reach or exercise the code that depends on DMA inputs. Second,
many embedded devices use DMA (roughly 25\% among the surveyed firmware,
see \S\ref{ap:surveyDMA}). In fact, DMA offers additional benefits to
embedded devices. For example, besides improving data transfer rates, DMA allows
processors to enter or stay in sleep or power-saving mode during data transfers,
which is desirable for power-constrained embedded devices. Third, DMA is the
only input channel used by certain peripherals and buses on embedded devices.
For instance, input from CAN (Controller Area Network) and USB is accessible to
firmware only via DMA.

\begin{figure}[]
    \includegraphics[width=0.4\textwidth]{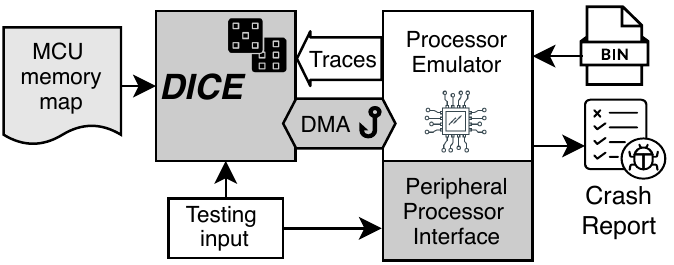}
    \centering
    \caption{\sysname design overview as a drop-in component of firmware analyzer}
    \label{fig:DICEoverview}
\end{figure}

\vspace{1em}
In this paper, we present \sysname (\sysnamelong), a drop-in component (Fig.~\ref{fig:DICEoverview}) for existing and
future dynamic firmware analyzers to recognize and manipulate DMA-based
peripheral input, and in turn, expand their analysis to cover firmware code,
states, and vulnerabilities dependent on DMA input. 

\sysname adopts a non-conventional approach to supporting firmware input during
analysis. Instead of aiming to emulate DMA controllers or DMA-enabled
peripherals, which is practically impossible due to the complexity and diversity
of these hardware components, we design \sysname to emulate the {\em DMA input
channels}, an implicit interface through which DMA input data flow to (or is
consumed by) firmware execution. These channels manifest as memory buffers which
peripherals directly write data to and firmware subsequently reads input from.
\sysname identifies \dins when the firmware programs the sources and destinations 
of \dtrans into the \dctr. The sources and destinations are addresses within 
specific ranges (such as RAM), and are written to a fixed region in memory, namely 
the memory-mapped registers of the \dctr. 
By identifying the creations and removals of such channels, \sysname maps the
sources and destinations of possible DMA data transfers. When firmware reads
from a memory buffer that belongs to a DMA data transfer destination, \sysname
detects it as a DMA input event and signals the analyzer for further actions
(e.g., supplying the read with fuzzer-generated input).

There are three main technical challenges that we tackled while pursuing this
approach. First, DMA input channels are created and removed on demand by
firmware during execution. \sysname needs to dynamically identify these channels
and monitor input events through them. Second, the high diversity in embedded hardware and 
software makes it difficult to develop a generic design for
\sysname that works on different types of architectures, peripherals, DMA controllers,
and firmware. Third, a large number of firmware analyzed in practice are in
binary form without source code or debug symbols. \sysname should not need source code for
identifying and emulating DMA input channels. Our design and implementation
addressed all these challenges. 

To evaluate the performance of \sysname and demonstrate the benefits that
it brings to existing firmware analyzers, we built two prototypes of \sysname, based 
on \syspp (for ARM Cortex-M) and a \mipspic 
(for MIPS M4K/M) \cite{PIC32emulator}, respectively. We performed 
unit tests or micro-benchmarking on 83 sample firmware, representing 11 different
MCU SoCs and covering ARM and MIPS architectures. 
\sysname detected 33 out of 37 \dins correctly, with 0 false positives. For the 22 channels
that firmware actually used during the tests, \sysname supplied inputs to 21 of
them without errors. We also conducted fuzzing tests on 7 real firmware and measured
the code and path coverages with and without \sysname enabled on \syspp. Thanks
to \sysname's automatically emulated \dins, the basic block coverage
increased by 30.4\% and the path coverage jumped remarkably by 79X.
Even with a fairly primitive memory sanitizer and a 48-hour fuzzing session,
\sysname detected 5 unique previously unknown bugs in the firmware.

The source code of \sysname, the integrations with
firmware analyzers and emulators, and all the firmware tested are publicly
available at \url{https://github.com/RiS3-Lab/DICE-DMA-Emulation}.

In sum, our work makes the following contributions: 

\begin{itemize}
    \item We study and advocate the importance of supporting DMA-based
    peripherals and input in dynamic firmware analysis; we identify the lack of
    DMA support as a common limitation for all existing firmware analyzers. 
    \item We present \sysname, which enables dynamic firmware analyzers to
    support DMA-based peripherals,  and recognize and manipulate DMA inputs for
    testing otherwise skipped or unreachable code/states in firmware.   
    \item When designing \sysname, we overcome the challenges posed by the
    dynamic nature of DMA, the diverse hardware and software of embedded
    devices, and the unavailability of firmware source code. 
    \item We implemented two prototypes of \sysname based on 
    \syspp and a MIPS \mipspic, respectively. Our evaluation shows 
    that \sysname: (1) achieved highly accurate \din identification and
    emulation, (2) helped \syspp significantly improve its analysis coverage and
    discover 5 new bugs in tested firmware. 
    \item We analyzed all the bugs discovered by \sysname and found all of them 
    are remotely exploitable. They have security consequences 
    such as information leakage, data corruption, and denial-of-service. 
    These bugs cannot be found by exiting firmware analyzers due to the lack of DMA support or emulation. 
\end{itemize}

\section{Motivation}
\label{motivation}

MCU-based embedded devices (or MCUs in short) integrate on a single chip a main
processor, RAM, Flash, and diverse peripherals. Their energy-efficiency and
sufficient computing power make them the ideal building blocks of IoT devices
and cyber-physical systems. The entire software stack on MCUs, referred to as
firmware, contains OS/system libraries, drivers, and application-level logic in a
monolithic form. 

Firmware, similar to other software, may contain programming errors or bugs that
can be exploited by attackers
\cite{ToyotaKillerFirmware,WifiPzero,WifiSelianiMarvellAvastar}. These security
bugs in firmware, although often similar in nature, can cause severe and unique
consequences because MCUs are widely used in mission-critical settings (e.g.,
industrial systems) and have direct physical outreach (e.g., controlling vehicle
movements). 

Unfortunately, bugs in MCU firmware are more difficult to detect than bugs in
conventional computer software, due to existing analysis tools' limited support
for highly diverse and heterogeneous MCU hardware. Specially, various kinds of
peripherals are used as the main communication channels through which firmware
communicates with other devices and interacts with the physical environment.
However, existing firmware analysis methods cannot fully model or emulate
peripherals, and thus, fail to trigger or reach a large portion of firmware code
during analysis, missing opportunities to detect bugs. 

\begin{figure}[t]
    \includegraphics[width=0.3\textwidth]{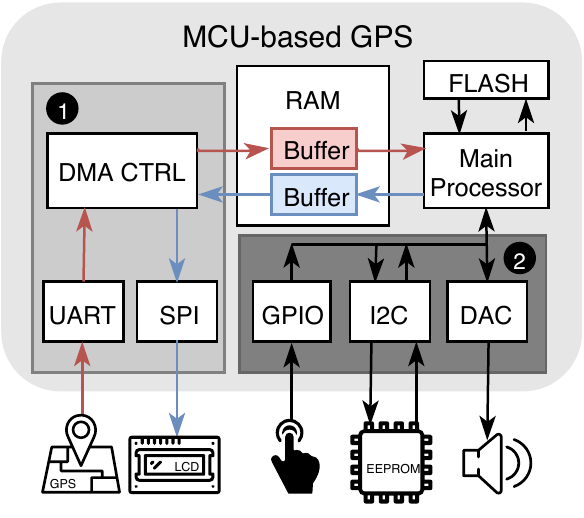}
    \centering
    \caption{A MCU-based GPS and its I/O flows: \textcircled{\small{1}} 
    DMA-based interfaces, 
    \textcircled{\small{2}} MMIO/interrupt-based interfaces.}
    \label{fig:NMEAGPS}
\end{figure}

Take a MCU-based GPS device as an example (Figure \ref{fig:NMEAGPS}). It uses
UART (universal asynchronous receiver-transmitter) to receive NMEA
\cite{NMEAgps} serialized messages from a GPS antenna. These messages are
copied to RAM via DMA. The firmware then parses the messages and computes the
location information, which is later copied via DMA to an LCD (Liquid Crystal Display) attached
to the SPI (Serial Peripheral Interface). This device also has other peripherals
for receiving inputs or delivering outputs. 

Without peripheral awareness or support, dynamic analysis of this GPS firmware
cannot reach most of the code because the firmware execution cannot receive any
input or even boot up the device.  Some recent work \cite{avatar,firmadyne,
p2im} addressed the peripheral dependence issue using different approaches.
However, they mostly focused on the simple peripherals that only use
memory-mapped I/O (e.g., those inside the box \textcircled{\small{2}} in Figure
\ref{fig:NMEAGPS}). \cite{inception} relies on manually identified DMA buffers
to partially support simple DMA-based I/O, and \cite{HALucinator} completely 
removes DMA through replacing HAL (hardware abstraction layer) functions with 
manually-written hooks. 

So far no existing work supports complex peripherals that use DMA to communicate
with firmware (e.g., those inside box \textcircled{\small{1}} in Figure \ref{fig:NMEAGPS}).
Therefore, firmware using DMA still cannot be fully analyzed and all their
bugs/vulnerabilities detected. According to our survey (\S\ref{ap:surveyDMA}), 
most nontrivial firmware use DMA for both performance and
energy-saving reasons. In these firmware, the majority of code cannot run or be
tested without DMA support.

\section{Background}
\label{sec:background}

\subsection{DMA on MCU Devices}
Direct Memory Access (DMA) is a widely used architectural feature that allows
peripherals to access the main system memory (RAM) without involving main
processors. The goal of DMA is to improve performance when transferring data
between RAM and peripherals. 
The introduction of DMA dates back to the 1960s on the DEC PDP-8 minicomputer~\cite{dec_pdp-8}. 
DMA is ubiquitously adopted by today's computers. 

DMA is also widely used in modern embedded devices powered by microcontrollers
(MCU). Similar to DMA on conventional computers, DMA on
MCU devices benefit from the performance improvement in data transfer between
RAM and peripherals. Unlike conventional computers, MCU devices use DMA not just
for performance reasons but also for saving power or energy. DMA allows large or
slow data transfers to take place while the main processor (i.e., a major power
consumer on embedded devices) is asleep or stays in the low-power mode.

From the programmers' perspective, DMA provides a standard interface that abstracts
away peripheral internals. Programmers can use the same interface exposed by DMA
controllers to exchange data with different peripherals. 
An example of this abstraction is the HAL library of ChibiOS \cite{chibiOS},
where the low-level drivers use DMA transparently through a uniform interface
to receive input from different peripheral.

We surveyed the availability of DMA on MCUs and the uses of DMA in
firmware. We analyzed 1,356 MCUs from a major vendor and 1,000 repositories from
Github targeting MCUs exclusively. Our analysis showed that 94.1\% of the modern
MCUs (32-bit architecture) are DMA capable, which demonstrates the ubiquitous
support for DMA on modern embedded devices. As for the firmware, 25.1\% of the
compilable or pre-compiled MCU repositories contain DMA related debugging
symbols, a strong indicator of firmware using DMA. The results underline the
importance and urgency of supporting DMA-capable peripherals and DMA-based input
in firmware analysis. We present the details of our survey in \S\ref{ap:surveyDMA}.

\subsection{DMA Workflow on ARM Cortex-M}
We choose ARM Cortex-M as the reference architecture for designing 
\sysname because it is the most common architecture used in modern MCU devices
and IoT. DMA works on this architecture in a similar way as it does on others, 
such as MIPS, which \sysname also supports. 
Below we describe a simplified DMA workflow and introduce the basic concepts,
which are necessary to understand the design of \sysname. 

\begin{figure}[t]
    \includegraphics[width=0.4\textwidth]{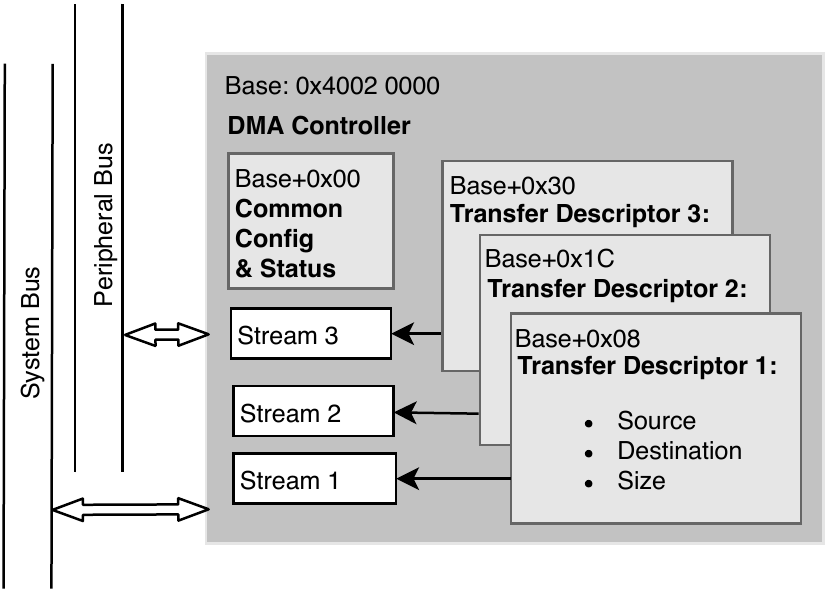}
    \centering
    \caption{DMA controller and detail of DMA \ddesrs, each using 20 bytes of memory span
    in the MMIO region (some fields of \ddesrs are omitted for simplicity).}
    \label{fig:DMAdescriptor}
\end{figure}

A {\em \dctr} plays the central role in the DMA workflow. It is an on-chip
peripheral that transfers data into memory on behalf of peripherals. A {\em
\dtran} is a single movement of data to memory. It starts with firmware creating
a {\em \ddesr}, specifying the transfer's source, destination, size, etc., and
writing this \ddesr to the \dctr, a step called DMA configuration. In addition,
firmware also specifies which {\em \dstr} (a physical data channel inside the
\dctr) should be used for the transfer. Figure \ref{fig:DMAdescriptor} shows a
\dctr with three \dstrs configured for three different \dtrans. After the
configuration finishes and the data becomes ready at the source, the \dctr performs
the transfer asynchronously without involving the firmware or the processor. 
The source and destination of a transfer are also referred to as
{\em \dptrs}.

A single MCU may have more than one \dctrs, each covering a set of connected
peripherals. A \dstr in a \dctr can only be used for supporting one \dtran at
any given time. High-speed on-chip peripherals, such as Ethernet, USB, and CAN,
may integrate a \dctr for higher throughput. Note that it is the
\dctr, not the source peripheral, that directly writes the data to memory.
\dctrs are the only peripherals that can directly access RAM thanks to the
special permission they have to access the system bus as master devices.

\subsection{Challenges of Supporting DMA in Firmware Analysis}

\point{Dynamic nature of DMA} 
The dynamic nature of DMA (\eg configurations, streams, and transfers) makes it
challenging to detect \dins or infer \dptrs. Some previous work relies on
manual and static identification of DMA buffers \cite{inception}. This 
strategy requires source code or reverse engineering 
of firmware. Despite the poor scalability, it can neither be complete nor 
accurate because it may miss dynamically created DMA buffers or mistakenly 
treat regular memory regions as DMA buffers. 
A reliable and practical approach should consider that DMA
transfers are created and performed dynamically on demand. It should not require
human efforts or rely purely on static inference.  

\point{MCU hardware and software diversity} 
The diversity of MCU firmware prevents the application of DMA models designed
for desktop OS, where a generic hardware abstraction layer (HAL) exists (e.g. for Linux
\cite{periscope}). For MCU firmware, there is no standard DMA interface or a
prevalent OS that provides a generic abstraction for various hardware
peripherals. 
Moreover, it is practically impossible to develop an emulator for
each \dctr and peripheral due to the high diversity of the hardware, as observed
by \cite{p2im,avatar,avatar2,firmadyne}. Therefore, it is necessary yet
challenging to design a single DMA emulation method that can work with a wide
range of firmware, architectures and peripherals.

\section{Survey on DMA Availability and Usage on MCUs}
\label{ap:surveyDMA}

We surveyed 1) the availability of \dctrs for MCUs on the complete product portfolio of a top MCU vendor, and 
2) DMA usage by firmware on a large collection of open-source repositories from GitHub.

\subsection{DMA Availability on MCUs}
We analyzed on March 2019 the MCU product portfolio (which documents all MCU models a vendor produced) of Microchip Semiconductors \cite{MICROCHIPproductportfolio}, a top MCU vendor in terms of market share \cite{TheMacleanReport2017}.
Our analysis included 1,356 MCUs from 32-bit (ARM Cortex-M, MIPS), 16-bit (PIC16), and 8-bit (PIC8 and AVR) architectures. We excluded the legacy architecture 8051.
We grouped MCUs into families using Microchip's quick reference guides \cite{MICROCHIP8BITS,MICROCHIP16BITS,MICROCHIP32BITS}.
MCUs within the same family, although different in memory size and packaging, use the same architecture and have mostly the same on-chip peripherals and DMA availability (as \dctr is also an on-chip peripheral). 
Therefore, grouping MCUs into families allow us to better analyze and describe DMA availability on those MCU families. 
Our analysis demonstrated that 94.1\% of modern 32-bit MCU families include one or multiple DMA controllers, or include DMA capable peripherals (e.g., USB, CAN, Ethernet).
On the other hand, only 11\% of 16-bit and 8-bit MCU families support DMA (Table~\ref{table:DMAcontrollersMCU}).

\begin{table}[t]
   \centering
    \begin{tabular}{l l l}
    \hline
    MCU architecture & DMA support & No DMA support \\
    \hline
     32-bit (ARM Cortex-M, MIPS) & 32 (94.1\%) & 2 (5.9\%) \\
     16-bit (PIC16) & 2 (11.1\%) & 16 (89.9\%) \\
     8-bit (PIC8, AVR) & 7 (11.3\%) & 55 (88.7\%) \\
   \hline
    \end{tabular}
    \caption{DMA availability on Microhip MCUs families (March 2019)}
    \label{table:DMAcontrollersMCU}
\end{table}

\subsection{DMA Usage by Firmware}
We collected 1,000 unique repositories from GitHub by searching keywords and topics related to microcontrollers, IoT, and DMA.
Our dataset only included repositories for ARM Cortex-M architecture, the most popular architecture for IoT devices and cyber-physical systems.
We found most repositories cannot be compiled because of missing building scripts (e.g., makefiles) or libraries, 
or unavailability of proprietary IDE (Integrated Development Environments) and toolchains. 
We were able to compile or directly download 350 ELF binaries from our dataset. 
We performed a basic static analysis on both source code and EFL files.

Our analysis shows that 920 out of 1,000 (92\%) repositories include DMA related
header files or DMA driver source code. However, only 88 out of 350 (25.1\%) 
ELF files contain DMA related debug symbols. 
This is because DMA header files and driver code are always distributed as part of SDKs, regardless of firmware usage. 
Therefore, we cannot use the inclusion of DMA header files or driver code as an indicator of DMA usage.
Instead, we use the appearance of DMA debug symbols in ELF files as the indicator for DMA usage by firmware, 
and estimate 25.1\% (88 out of 350) firmware use DMA.

We also observed that most firmware that use DMA are for battery-powered devices (e.g., drones, handhelds, smart watches), 
or require high data communication throughput (e.g., DSPs and LCDs). %
We admit that usage of DMA is an architectural and/or design decision made by firmware developers for the specific application.

\section{System Design}
\label{sec:design}

\sysname provides dynamic firmware analyzers the ability to recognize and handle
DMA-based input from peripherals, thus allowing firmware using DMA to be
analyzed and firmware code dependent on DMA input to be executed and tested,
which is previously impossible. \sysname meets the following design goals that
we set: 
\begin{itemize}
    \item {\bf Hardware independence:} \sysname should not rely on actual
    hardware peripherals. It should be generic to support a wide range of
    architectures, peripherals, and DMA controllers used in embedded devices.    
    \item {\bf Firmware compatibility:} \sysname should be compatible with all
    possible ways that firmware may use DMA as input channels, regardless of how
    DMA is configured or data is consumed.  
    \item {\bf Dynamic DMA:} \sysname should fully consider the dynamic nature of
    DMA and be able to capture DMA input events through dynamically allocated
    memory regions.  
    \item {\bf No source code:} \sysname should not require source code or debug symbols of firmware.  
    \item {\bf Integration with analyzers:} \sysname should not need hardware or
    software capabilities that common firmware analyzers do not have.
    Integrating \sysname should not require major changes to existing analyzers.
    
\end{itemize}

\sysname achieves the design goals thanks to a novel
approach to supplying DMA input to the firmware.  The approach is inspired by our
observation on the generic patterns that firmware follows when performing DMA
configurations and data transfers. These patterns are observed across different
firmware on various embedded devices using distinct DMA controllers and
peripherals. In fact, these patterns reflect the de facto protocol used by
embedded firmware and peripherals when exchanging data via DMA. By detecting
these patterns and intercepting the DMA configuration and data transfer events,
\sysname monitors \dins as they are created, used, and disposed. As a result,
\sysname can capture (and respond to) all DMA input events, which inevitably go
through the monitored interfaces. 

Next, we explain the abstract notion of \dins and discuss the DMA configuration
and data transfer patterns that \sysname uses to identify \dins.

\subsection{DMA Input Channels}
{\em DMA input channel} is an abstract notion we formulated, on which the core
idea of \sysname is derived. Such channels can be viewed as the conceptual
bridges through which firmware and peripherals exchange data via DMA. They
manifest as dynamically allocated memory buffers that firmware and peripherals
agree upon for transferring data. A \dctr serves as the proxy for a peripheral
to write data in the DMA memory buffers without involving the main processor. The
data is then read from the memory buffer by the firmware as input from the
peripheral. Therefore, if all \dins (or their manifestations, i.e., memory
buffers used as DMA data exchanges) can be recognized upon their creation,
access, and disposal, all DMA data transfers can be monitored and interposed,
which allows \sysname to supply DMA input to the firmware execution without using
actual peripheral hardware or understanding the inner workings of peripherals
or DMA controllers. 
\textit{\dout} follows the same definition with \textit{\din}, but with the opposite 
transfer direction (i.e., data is transferred to peripherals via DMA).

\begin{table}[t]
    \centering
     \begin{tabular}{l l c l} 
     \hline
     \textbf{Source} & \textbf{Destination} & \textbf{Valid} &\textbf{Type} \\
     \hline
        Peripheral  &RAM        &Yes & \din  \\
        Peripheral  &Peripheral &Yes & \dout   \\
        Peripheral  &Flash      &No  & N/A  \\
        RAM         &Peripheral &Yes & \dout   \\
        RAM         &RAM        &Yes & \din  \\
        RAM         &Flash      &No  & N/A  \\
        Flash       &RAM        &Yes & \din  \\
        Flash       &Peripheral &Yes & \dout   \\
        Flash       &Flash      &No  & N/A  \\
     \hline
     \end{tabular}
     \caption{Possible combinations of sources and destinations for DMA input and output channels}
     \label{table:validpointer}
\end{table}

Table \ref{table:validpointer} shows all possible combinations of sources and
destinations for DMA input and output channels. Those with Flash as the destinations are invalid because
DMA is only meant for transferring data to memory, including RAM and
memory-mapped peripheral regions. Among the valid combinations, we focus on
those with RAM as the destination because they are \dins that
can directly influence the execution and analysis of firmware. 
\sysname does not handle \douts which do not directly influence firmware 
execution. Unless otherwise noted, we refer to \dtrans that go through \dins simply as \dtrans. 

\begin{figure}[t]
    \includegraphics[height=0.45\textwidth]{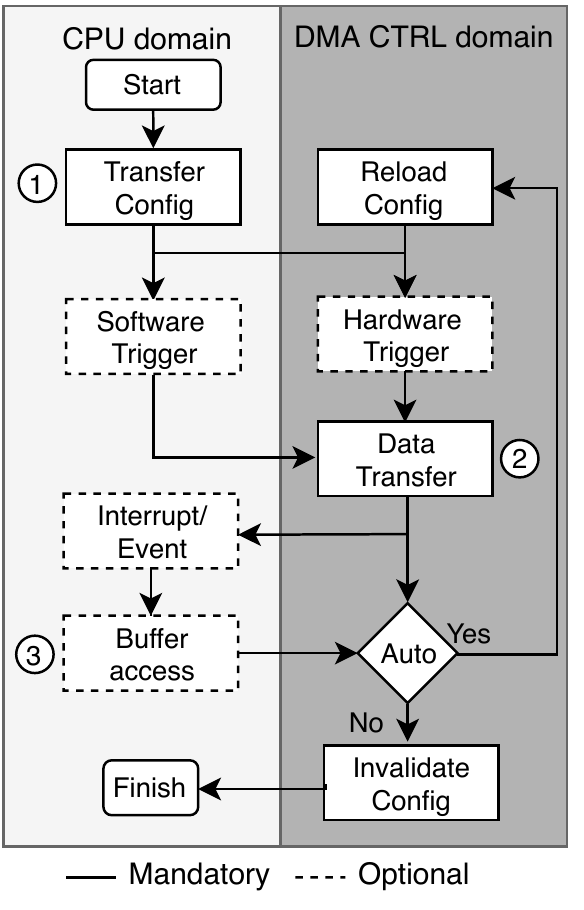}
    \centering
    \caption{Life-cycle of a DMA transfer}
    \label{fig:DMALifeCycle}
\end{figure}

Each \dtran takes three steps. First, firmware establishes
the DMA input channel by sending a \ddesr to the \dctr as part of the transfer
configuration (\textcircled{\small{1}} in Fig.~\ref{fig:DMALifeCycle}). The
\ddesr specifies, among other things, which physical \dstr inside the \dctr
should be used for transferring the data, along with the source and destination
addresses as well as the transfer size. The source address points to the
to-be-transferred data stored in peripheral registers, Flash, or RAM. The
destination points to the memory buffer where the firmware expects the input
data to be transferred. We refer to a source or destination address as a \dptr.
Second, when the input data becomes available at the source, as indicated by an
interrupt, the \dctr copies the data from the source to the destination. Note
that the actual data transfer and direct memory access are performed by the
\dctr on behalf of peripherals (\textcircled{\small{2}} in Fig.~\ref{fig:DMALifeCycle}). 
Third, after finishing the data transfer, the \dctr
signals the firmware and closes the \dstr. It tears down the current DMA input
channel and leaves the input data in the memory buffer for the firmware to use
(\textcircled{\small{3}} in Fig.~\ref{fig:DMALifeCycle}). 

A \ddesr is valid only for one \dtran, which goes through the DMA input channel
identified by the \ddesr. Firmware may sometimes use the auto-reload feature of
\dctrs, whereby a \ddesr is valid for multiple consecutive \dtrans until
reconfigured. 

The lifespan of a \din starts when \dctr receives the \ddesr and the \dstr is
configured. It terminates when the data transfer finishes. A channel's exit end
is attached to a dynamically allocated memory buffer (i.e., the destination).
This per-transfer and highly dynamic nature of \dins makes it challenging to
detect and monitor them. For the same reason, previously proposed techniques for
detecting DMA input using static heuristics are inaccurate and incomplete.

\subsection{Capturing Stream Configurations}
\label{sec:design_streamconf}

\sysname dynamically captures \dstr configuration events to extract the
information needed for identifying \dins, such as destinations and sizes.
However, stream configurations are not directly visible  by firmware emulators
or analyzers due to the semantic gap---such a configuration (i.e., firmware
writing a \ddesr to memory-mapped registers), in the eyes of an emulator, looks
the same as a regular memory write by firmware.

We observed a fairly distinct pattern followed by \dstr configurations: 
writing some values within a specific range to a fixed region in memory. This
pattern echos the essential operation performed in each stream
configuration---writing the source and destination \dptrs 
to the \dctr registers. These peripheral registers are always mapped in the MMIO region 
(0x40000000–0x5fffffff). 
\dtrans through \dins move data from 
peripheral/Flash/RAM to RAM. Therefore, the value of source \dptrs must be in 
the range of 0x40000000–0x5fffffff (for peripheral MMIO), 0x20000000-0x20004fff 
(for RAM), or 0x8000000-0x801ffff (for Flash) 
\footnote{The RAM and Flash address ranges are taken from the STM32F103 MCU as an example. 
Different MCU may use slightly different ranges for RAM and Flash, which are specified in their 
data-sheets. When data-sheets are not available, \sysname uses the largest ranges allowed by 
the architecture (a 512MB region) as the RAM/Flash region.}. 
Similarly, destination pointers must point to the RAM region. 
These regions are not very big in size (2kB to 512kB), 
which means the value range of \dptrs is not very wide. 
\sysname uses this pattern (\ie two writes of pointer values to consecutive locations in
the MMIO region) to detect \dstr configurations.

We also observed that a \dctr can support multiple \dstrs
(Figure~\ref{fig:DMAdescriptor}). The source and destination 
\dptrs of a particular stream are recorded in two consecutive MMIO registers in the \dctr. 
The write operations to these registers are 32-bit in width and 4-byte aligned in address.
This pattern allows \sysname to identify multiple stream configurations
on the same \dctr, and more importantly, filter out  
pointer-like values that are written to the MMIO region yet not \dptrs (\ie the write operations are sparse, unaligned or in a different data width). 

Based on our tests on real firmware and devices of various kinds
(\S\ref{sec:unit_test}), this pattern reliably indicates stream configurations
and is never seen in other types of memory write operations. It means that,
empirically, firmware writes two pointer/address values to two consecutive MMIO
registers only for the purpose of \dstr configuration. %

\sysname looks for the stream configuration pattern while it monitors memory
writes by firmware during execution. When one is observed, \sysname captures the
\dstr configuration and extracts from it the \dptrs. Then \sysname needs to 
identify the direction of the \dtran, because when \sysname identifies two 
\dptrs that point to, for example, peripheral and RAM regions, it does not know 
whether the \dtran is from peripheral to RAM, or the opposite.

\sysname determines the transfer direction by monitoring memory accesses made by
the firmware through the \dptrs. If the firmware reads from the RAM address referenced by the \dptr, the \dtran is from peripheral to RAM (a 
\din that we are interested in). If the firmware writes to the RAM address referenced by the \dptr, the \dtran is from RAM to 
peripheral. This is determination of transfer direction is intuitive. For incoming DMA data, firmware needs to read the data after it is transferred into RAM from a peripheral. For outgoing data, firmware needs to write it into 
RAM before DMA transfers it to a peripheral. 

\sysname captures a \dstr configuration transferring data to RAM, and finds its destination 
address (i.e., the beginning of the memory buffer for receiving the 
current \dtran).  A newly captured \dstr configuration marks the 
establishment of a \din for an upcoming \dtran. Its destination address 
locates the memory buffer that the firmware will read 
the DMA input from. \sysname keeps track of all active \dins and their
destination addresses. 

\point{Pattern variations} We encountered one variant of the
stream configuration pattern described above, namely,  
multiple destination addresses are specified in one
configuration. This occurs when a \dtran operates in the circular mode
\cite{STM32F4}. Although slightly different from normal stream configurations,
this variation still obeys the pattern that \sysname uses to capture stream
configurations. It simply uses one (or more) extra destination \dptr (\eg three 
pointers written to three consecutive MMIO peripheral registers).

\point{Pattern limitation} We identified two limitations of the pattern for 
capturing \dstr configuration. First, the pattern assumes that firmware always writes
\ddesrs, including the \dptrs, to the memory-mapped peripheral region that
corresponds to \dctr registers. Although this is true for most firmware and MCU
devices, we are aware of some rare cases where firmware writes \ddesrs to RAM,
rather than \dctr registers. This type of stream configuration is 
used only in some high-end SoCs, which resembles the desktop architectures.  
In this case, \dctrs fetch \ddesrs from RAM in ways unique to the individual 
\dctrs. We did not encounter such cases in our experiments. 
\sysname cannot capture the stream configurations performed 
this way. 

Second, some \dctr models (e.g., NRF52832 easyDMA \cite{nRF52832datasheet})
require only the destination address, not the source-destination pair, when
firmware configures a DMA stream (using an implicit source address). As \sysname
needs to observe both the source and destination \dptrs in order to identify a
\dstr configuration, \sysname cannot identify destination-only stream
configurations. However, based on our evaluation, only 2 \dctr
models support this type of configuration. \sysname missed only 7 out of 52 \dstr
configurations due to this limitation. 
We discuss the details in \S\ref{sec:evaluation}.

\subsection{Responding to DMA Data Read}
\label{sec:design_read}

After \sysname captures a \dstr configuration and finds the destination address, 
it places an access hook on the destination address, which allows 
\sysname to identify and respond to firmware's read from the corresponding DMA memory buffer. 
However, this DMA read identification and response process is not as straightforward 
as it may sound, due to two technical challenges: 
the unknown buffer size, and the dynamic termination of \dins. 

\point{Unknown buffer size} 
Although \sysname can reliably capture every \dstr configuration and extract the
destination address, it cannot accurately find the transfer size or the buffer
size from a captured configuration event. This is because transfer sizes may
take a wide range of values, unlike destination addresses, whose values are
bounded by the valid DMA memory regions and therefore fairly distinguishable.

\sysname needs to know where each DMA buffer ends in order to determine if a
memory read falls in such a buffer. 
An intuitive solution is to extract buffer size from the debug symbols generated
by compiler. At \dstr configuration, debug symbols are looked up to figure out 
which buffer is allocated at the destination address and what the buffer size is. 
However, debug symbols are absent in MCU firmware, which are stripped binary 
blobs containing only code and data. Moreover, this approach cannot identify the
size of dynamically allocated buffers that are widely used in \dtrans. 

Instead, \sysname adopts an approach supporting dynamically created DMA buffers 
without using debug symbols. It dynamically 
infers the bounds of DMA buffers by observing firmware's access. The inference
leverages the fact that firmware typically reads a DMA buffer consecutively in
space (from the beginning to the end, but not necessarily consecutive in time).
Although in theory firmware may not start reading DMA buffers from the
beginning, we did not observe such a case in our experiments on real firmware.
When input data comes from a peripheral that uses a different endianness than
the MCU's, firmware may start reading the buffer several bytes after the
beginning. \sysname considers and handles such cases.

For each firmware access to a detected DMA buffer, \sysname calculates a
\textit{span} (\ie the possible extent to which this DMA buffer may extend
beyond the currently known boundary). The \textit{span} size is set to twice of
the data width of the memory read operation (LOAD). For instance, the span for a
buffer accessed via a 32-bit LOAD will have the size of 64 bits (8 bytes).
Having a span for each buffer access allows \sysname to incrementally infer the
buffer size and recognize the endianness conversion that the firmware may
perform. For example, after multiple two-byte inputs are transferred from
big-endian peripherals by DMA, firmware running on a little-endian CPU will read
the DMA buffer in 1-byte data width to convert the endianness. As byte 1 is read
before byte 0, having a \text{span} twice of the data width allows \sysname to
catch this behavior. \sysname monitors memory read operations while the \din is
in use. When a memory read falls in the span, \sysname expands the detected DMA
buffer to include the read address. This process is described in Algorithm
\ref{algo:1}.   The dynamic and incremental expansion of detected DMA buffers
allows \sysname to identify and handle firmware's DMA read while continuously
inferring the true buffer size, especially for dynamically allocated DMA
buffers. 

\begin{algorithm}
\scriptsize
\caption{DMA buffer size inference}
\label{algo:1}
\begin{algorithmic} 
    \STATE $PerceivedSize \leftarrow 0$
    \WHILE{ \din  is valid }
    \IF{memory is read}      
            \STATE $Span.Size \leftarrow 2*Read.DataWidth$
            \STATE $Span.Base \leftarrow (Buffer.Base + PerceivedSize)$
            \IF{Read falls in  Span}
                \STATE $PerceivedSize \leftarrow (Read.addr-Buffer.Base) + Read.DataWidth$
            \ENDIF
        \ENDIF
    \ENDWHILE

\end{algorithmic}
\end{algorithm}

\point{Dynamic channel terminations}
\dins are per DMA transfers and not permanent. A channel is created when the
underlying \dstr is configured and terminated when the stream is invalidated or
used for a different transfer. \sysname tracks the life cycles of \dins and
detects dynamic channel terminations. This is necessary because stale channels,
if not recognized, can cause \sysname and the firmware analysis to mistakenly
treat regular memory access as DMA input events and thus corrupt firmware
execution. It is worth noting that the static technique used by the previous
works for identifying DMA buffers suffer from this issue. 

\sysname watches for two types of signals that indicate channel terminations.
First, if a newly captured \dstr configuration references the same \dstr as a
previously captured one did, the stream is now reconfigured to support a
different \dtran, and therefore, the previous \din is now terminated. Second, if
firmware writes to a memory buffer that corresponds to a \din, the channel is
implicitly terminated and the buffer is no longer used for receiving DMA input.
When a channel termination is detected, \sysname removes the access hook on the
buffer. As a result, \sysname no longer treats it as a DMA buffer or supplies
input to it. The previously generated input data still remains in the buffer,
which firmware may continue using.

\section{Implementation}
\label{sec:implementation}

\begin{figure*}[t]
    \includegraphics[height=3.7cm]{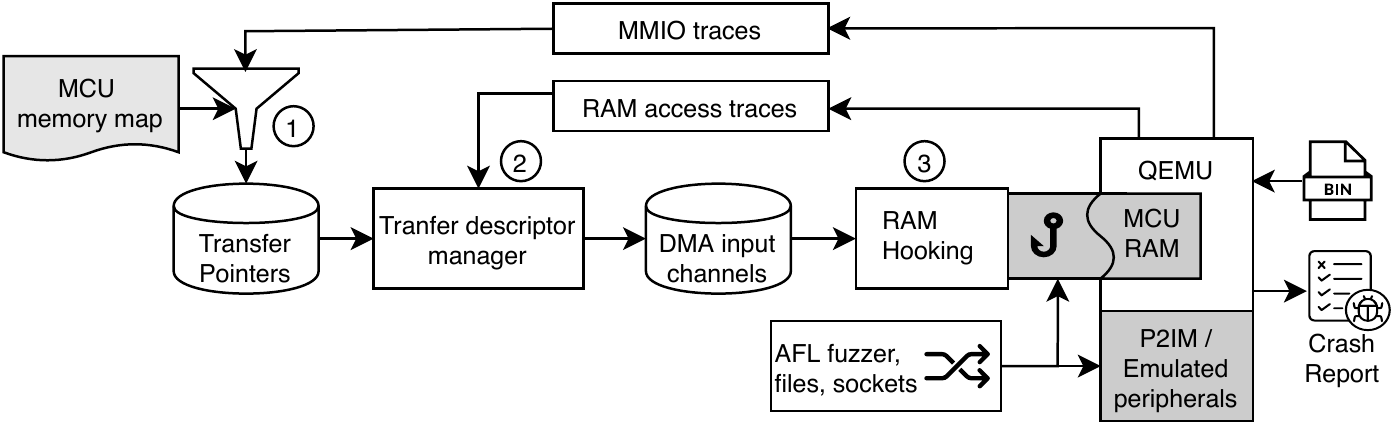}
    \centering
    \caption{\sysname implementation (\syspp version). \textcircled{\small{1}} Identification of DMA \dptrs. \textcircled{\small{2}} \ddesr management. \textcircled{\small{3}} Transfer size and DMA hooks. }
    \label{fig:DMAoverview}
\end{figure*}

We implemented two prototypes of \sysname as drop-in components
on top of the \syspp framework (for ARM Cortex-M architecture) and a \mipspic (for MIPS M4K/M-Class architecture) \cite{PIC32emulator}.

\sysname leverages the tracing and hooking capabilities of QEMU, which is the 
processor emulator used in \syspp and the \mipspic. It is worth noting that 
these emulators/analyzers do not provide any DMA capabilities as part of their original design.
Also, \sysname does not rely on any internals of these systems, 
making \sysname portable to other firmware analyzers.

Although the implementation of \sysname in the two prototypes are virtually the 
same, they could not be integrated under a single
implementation because these systems are based on different and heavily modified 
QEMU forks. 
Also, each emulator has a different set of capabilities. \syspp can run a broader range of 
firmware  and perform fuzz testing thanks to its automatic modeling of 
processor-peripheral interfaces and integration with AFL \cite{afl}. The \mipspic can only 
run a small set of firmware because it emulates a limited number of peripherals. 
We consider extending \syspp to support MIPS M4K/M-Class architecture, or port 
\syspp to the \mipspic out of the scope of this paper.

On the \mipspic, any access to non-emulated peripherals will halt the emulator, which 
significantly limits firmware execution. To avoid halting the firmware, we 
created default memory-mapped register handlers in the emulator. These handlers 
do not implement any meaningful peripheral functionalities. They treat registers
as memory, i.e, they record the value that is written to a register, and return 
it when the register is read. 
Besides, we implemented a round-robin strategy (inspired by \syspp 
design) for triggering DMA-related interrupts and invoking the associated ISR 
(interrupt service routine). This allows the firmware to read DMA buffers.

The DICE implementation on \syspp includes 1,237 lines of C code added to QEMU. The implementation on \mipspic has 1,620 lines of C code. 
Despite the different target architectures (ARM and MIPS), 
both implementations are fairly similar and the description below applies to both.  
The heuristic to capture \dstr configurations is implemented in QEMU's
\texttt{unassigned\_mem\_write} function, which manages MMIO operations related to 
the peripheral memory area (\textcircled{\small{1}} in Fig.~\ref{fig:DMAoverview}). 
This function has access to traces of the MMIO operations. These traces 
include information about the operation type (read/write), value, address and 
data width (8, 16 or 32 bits) of the MMIO.
The \ddesr manager (\textcircled{\small{2}} in Fig.~\ref{fig:DMAoverview}) is a software component 
that tracks and orchestrates the life cycles of \dins. We implemented this 
component and data structures in the \texttt{helper\_le\_ld\_name} function. This function is 
defined in the softmmu\_template.h file and has access to traces of any memory read in all
memory areas of MCU. We modified QEMU to invoke this function on every LOAD
instruction. This method is described as the ``slow path" for 
memory access, according to QEMU's documentation \cite{qemu_sp}. 

The \texttt{helper\_le\_ld\_name} function allows \sysname
to place memory hooks on-the-fly (\textcircled{\small{3}} in Fig.~\ref{fig:DMAoverview}),
according to the captured \dins and accesses of firmware to the destination address.
Also, \texttt{helper\_le\_ld\_name} provides the effective data width (8, 16 and 32-bit)
of the LOAD instruction emulated by QEMU. This information is used to adjust
the \textit{span} that is used to identify the buffer size. 

The \sysname implementation on \syspp supports fuzzing. It reuses the underlying 
AFL engine and the TriForce \cite{triforceafl} QEMU 
extensions of \syspp.  \sysname supports
files, network sockets, and other input methods that a firmware analyzer requires 
to provide input to firmware. This architecture allows \sysname to be added as a drop-in
component to other firmware analyzers and enhance it with the automatic
emulation and manipulation of \dins.

\section{Evaluation}
\label{sec:evaluation}

We evaluated \sysname from three different angles: 1) whether it can accurately
identify \dins on firmware that run on different architectures, MCUs and OSes; 
2) how much its runtime overhead is; 3) whether it can support fuzz-testing on real firmware
that uses DMA, and more importantly, find bugs that cannot be found by existing
dynamic firmware analyzers. 

To verify 1) and 2), we performed unit tests and micro-benchmarks on sample
firmware in \S\ref{sec:unit_test}. As for 3), we fuzz-tested 7 real-world
firmware with \sysname (integrated with \syspp) and found 5 previously unknown
bugs in \S\ref{sec:fuzzing}. We also discussed our insights into DMA emulation.

All experiments were conducted on a
dual-core Intel Core i5-7260U CPU @ 2.20GHz, 8 GB of RAM, and a fresh
installation of Ubuntu 18.04 LTS. We will release all the firmware images after
the paper is published.

\subsection{Unit Tests on Sample Firmware}
\label{sec:unit_test}
We conducted this experiment to show that \sysname can accurately identify 
\dins on different architectures, MCUs and OSes (i.e., \sysname is accurate,
hardware-independent and OS-agnostic). We collected a set of 83 sample firmware 
from the official MCU SDKs and open-source repositories. These sample firmware 
are developed by MCU vendors or open-source contributors and serve as templates for 
firmware developers. The sample firmware are suitable for our micro-benchmarking 
because each implements a self-contained logic and they collectively cover: 
1) different architectures, vendors, MCUs, and \dctrs, 2) different
OSes/system libraries, 3) different combinations of DMA sources and
destinations.

\point{Experiment Setup} 
As shown in Table \ref{table:MCUs}, the 83 sample firmware cover 2 architectures, 
11 different MCUs from 5 major vendors, and 9 unique \dctr models.

\begin{table}[t]
     \centering
      \begin{tabular}{llll@{}}
      \hline
      \textbf{MCU} & \textbf{Architecture} & \textbf{DMA cntlr.} &  \textbf{Vendor} \\
      && \textbf{model} &\\
      \hline
      NRF52832      &  ARM Cortex-M4  &   a     &  \multirow{2}{2cm}{Nordic\\Semiconductors}  \\
      NRF51822      &  ARM Cortex-M0  &   a                           &                                             \\   
      \hline
      NUC123        &  ARM Cortex-M0  &   b                          &    Nuvoton \\
      \hline
      LPC1837       &  ARM Cortex-M3  &   c                          & \multirow{2}{1.8cm}{NXP} \\
      MK64F         &  ARM Cortex-M4  &   d                          &                           \\
      \hline
      SAM3X         &  ARM Cortex-M3  &   e,f                          &  \multirow{3}{1.8cm}{Microchip/\\Atmel}\\
      PIC32MX795    &  MIPS M4K   &   g       &    \\
      PIC32MZ2048   &  MIPS M-class &  g                          &  \\  
      \hline
      STM32F103     & ARM Cortex-M3    & h      &    \multirow{3}{1.8cm}{ST \\Microelectronics} \\
      STM32L152     & ARM Cortex-M3    & h                           & \\
      STM32F429     & ARM Cortex-M4    &    i                        & \\
      \hline    
     \end{tabular}
     \caption{Architectures, MCUs, \dctr models and vendors covered by the 83 
     sample firmware. Some MCUs share the same \dctr model. SAM3X has two different \dctr models. For brevity, we use the letters (a-i) to differentiate the controller models.}
     \label{table:MCUs}
    \end{table}

The sample firmware are based on the real-time OSes (RTOS) 
or system libraries chosen by the vendors (SDKs), including BSD, Arduino, Mynewt, NuttX, Riot OS and 
ChibiOS (Table~\ref{table:FirmwareUnitTest}). This set of firmware include not only 
those that use all types of \dins, but also those that do not use DMA (non-DMA-enabled firmware). 
The non-DMA-enabled firmware include the whole unit test suite used in \cite{p2im} and 
2 ports of BSD for MCUs. 
Including both DMA-enabled and non-DMA-enabled firmware allows us to comprehensively evaluate
\sysname, in terms of its accuracy and compatibility.

\begin{table*}[t]
     \centering
      \begin{tabular}{p{3cm} p{2cm} p{2cm}   p{7.2cm} }    
      \hline
      \multicolumn{4}{c}{\textbf{ARM Cortex-M0/M3/M4 DMA-Enabled Firmware}}   \\
     
      \textbf{Firmware} & \textbf{MCU} & \textbf{OS/SDK} &  \textbf{Source Code} \\
      \hline
      ADC PDC * &  SAM3x  &  Arduino  &   \url{http://nicecircuits.com/playing-with-analog-to-digital-converter-on-arduino-due/} \\
      \hline
      SPI DMAC Slave * & \multirow{2}{2cm}{SAM3x} &  \multirow{2}{2cm}{Atmel ASF}  &   \multirow{2}{6cm}{\url{https://asf.microchip.com/docs/latest/sam.drivers.spi.spi_dmac_slave_example.sam3x_ek/html/index.html}} \\
      USART DMAC      &     &   &        \\
     \hline
      ADC slider          &  STM32F103  &     \multirow{7}{2cm}{ChibiOS}  &   \multirow{7}{6cm}{\url{https://osdn.net/projects/chibios/downloads/70739/ChibiOS_19.1.0.7z/}}\\
      I2C accelerometer   &  STM32F103                                  & &   \\
      SPI                 &  STM32F103                                  & &   \\
      UART                &  STM32F103                                  & &   \\
      ADC slider          &  STM32F429                                  & &   \\
      SPI                 &  STM32L152                                  & &   \\
      UART                &  STM32L152                                  & &   \\
      \hline
      ADC SW DMA *         &  \multirow{14}{2cm}{ STM32F103 }  &  \multirow{14}{2cm}{STM32CubeF1}    & \multirow{13}{6cm}{\url{https://www.st.com/en/embedded-software/stm32cubef1.html}} \\
      ADC Timer DMA *      & &                                 &   \\
      I2C DMA IT          & &                                 &   \\
      I2C DMA Adv IT          & &                                 &   \\
      I2C TxRx DMA *      & &                                 &   \\
      SPI Half DMA  *      & &                                 &   \\
      SPI Half DMA Init   & &                                 &   \\
      SPI Full DMA *      & &                                 &   \\
      USART TxRx DMA *    & &                                 &   \\
      USART Full DMA *    & &                                 &   \\
      UART H.Term. DMA    & &                                 &   \\
      UART 2Boards DMA    & &                                 &   \\
      SPI Full EX. DMA *   & &                                 &   \\
      I2C 2Boards DMA     & &                                 &   \\
      \hline
      PDMA M-M  &  LPC1837  &  LPC Open    & \url{https://www.nxp.com/downloads/en/software/lpcopen_3_02_lpcxpresso_mcb1857.zip} \\
      \hline
  
      \multirow{2}{2cm}{Serial console}  &  NRF52832  &  \multirow{2}{2cm}{Mynewt}     & \multirow{2}{6cm}{\url{https://mynewt.apache.org/download/} }\\
                                         &  NRF51822  &                                & \\
      \hline
      SPI slave  &  NRF51822  & \multirow{4}{2cm}{Nordic SDK}    & \multirow{4}{6cm}{\url{https://www.nordicsemi.com/Software-and-tools/Software/nRF5-SDK/Download}} \\
      SPI master  &  NRF52832   &                                &  \\
      SPI slave  &  NRF52832    &                                &  \\
      Serial DMA  &  NRF52832   &                                &  \\
      \hline
      PDMA USART  & \multirow{2}{2cm}{NUC123}  & \multirow{2}{2cm}{OpenNuvoton}    & \multirow{2}{6cm}{\url{https://github.com/OpenNuvoton/NUC123BSP.git}} \\
      PDMA M-M  &    &     &  \\
      \hhline{====}
      \multicolumn{4}{c}{\textbf{ARM Cortex-M3/M4 Non-DMA-Enabled Firmware}}   \\
      \textbf{Firmware} & \textbf{MCU} & \textbf{OS/SDK} &  \textbf{Source Code} \\
      \hline
      \multirow{3}{3cm}{\syspp unit test suite \\ (44 firmware images) \cite{p2im}}& STM32F103 & \multirow{3}{2cm}{Arduino, Riot OS, NuttX} & \multirow{3}{6cm}{ \url{https://github.com/RiS3-Lab/p2im-unit_tests}} \\
                             & SAM3x     &                                             &    \\
                             & MK64F     &                                             &    \\
     \hhline{====}
      \multicolumn{4}{c}{\textbf{MIPS M4K/M-class DMA-Enabled Firmware}}   \\
      \textbf{Firmware} & \textbf{MCU} & \textbf{OS/SDK} &  \textbf{Source Code} \\
     \hline
      PIC32MX\_UART               & PIC32MX795   &  \multirow{4}{2cm}{Microchip \\ Harmony v3} &  \multirow{4}{6cm}{\url{https://microchipdeveloper.com/harmony3:pic32mx470-getting-started-training-module}\\\url{https://microchipdeveloper.com/harmony3:pic32mzef-getting-started-training-module}} \\        
      PIC32MX\_test               & PIC32MX795   &                                              &    \\
      PIC32MZ\_UART               & PIC32MZ2048    &                                              &    \\    
      PIC32MZ\_ef\_curiosity      & PIC32MZ2048    &                                              &    \\   
      \hhline{====}
      \multicolumn{4}{c}{\textbf{MIPS M4K/M-class Non-DMA-Enabled Firmware}}   \\
      \textbf{Firmware} & \textbf{MCU} & \textbf{OS/SDK} &  \textbf{Source Code} \\
      \hline
      PIC32MX\_RetroBSD           & PIC32MX795     &  RetroBSD                                    &  \url{http://retrobsd.org/wiki/doku.php/start} \\ 
      PIC32MZ-BSD-Lite            & PIC32MZ2048    &  LiteBSD                                     &  \url{https://github.com/sergev/LiteBSD/wiki}   \\
     \hline
  \end{tabular}
      \caption{Sample firmware tested in unit tests. Firmware marked with * were also used in performance tests.}
      \label{table:FirmwareUnitTest}
     \end{table*}
  \label{sec:poc}

\begin{figure}[t]
    \includegraphics[width=8cm]{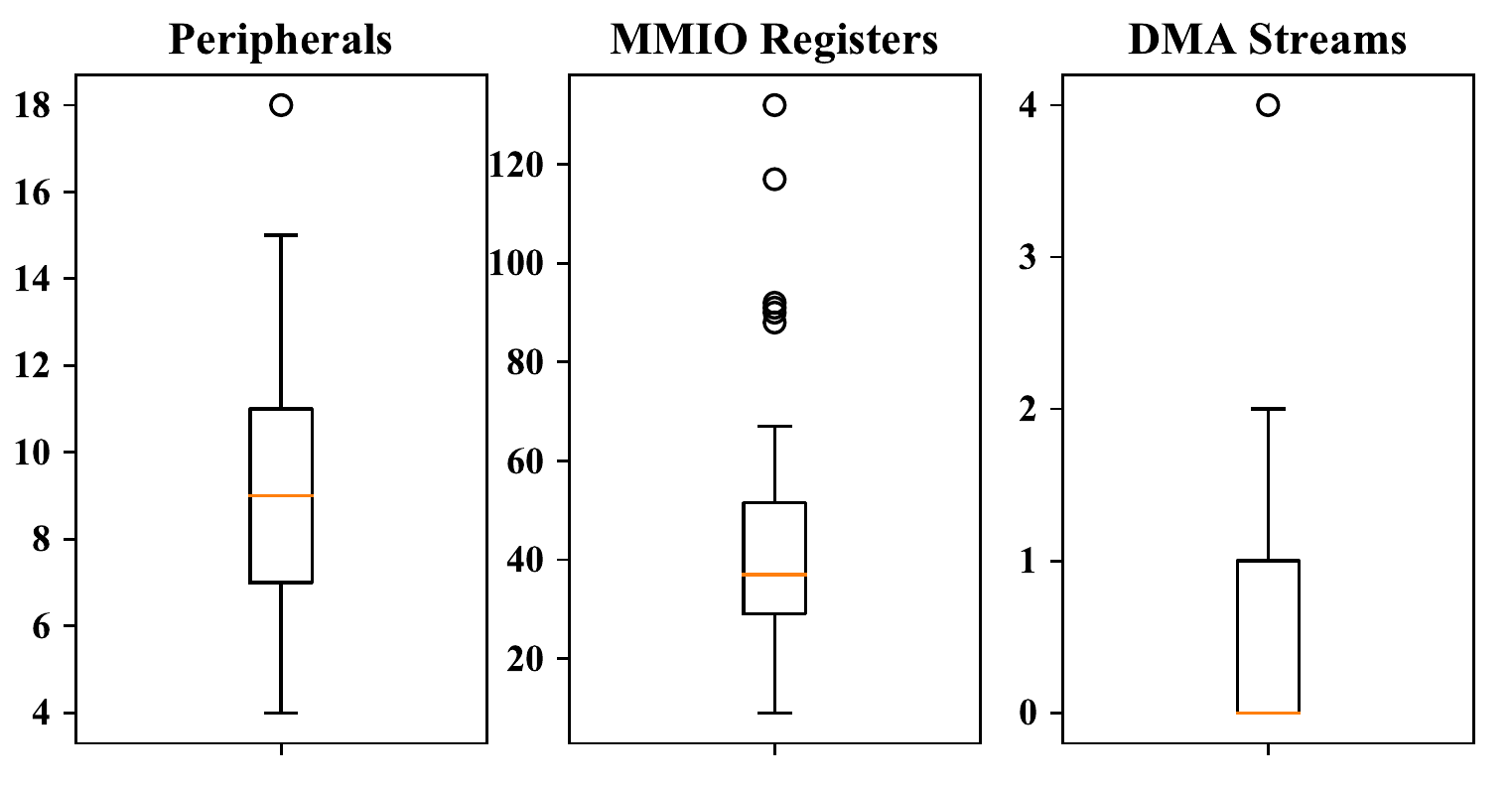}
    \centering
    \caption{Distributions of numbers of peripherals, MMIO registers and DMA stream configurations used in sample firmware (outliers marked by $\circ$). 
    All firmware access multiple and different peripherals, which shows their complexity and diversity.}
    \label{fig:statsUnitTest}
\end{figure}

All the sample firmware include the essential routines for running on real
devices, such as OS initialization and peripheral (including \dctr)
configuration and operation. As shown in Figure \ref{fig:statsUnitTest}, each
firmware accesses multiple peripherals (ranging from 4 to 18) and registers
(ranging from 9 to 132). Each firmware configures up to 4 DMA streams
simultaneously.

We run each unmodified firmware binary with the \sysname prototypes supporting 
ARM and MIPS architectures and collected debug output 
from QEMU for evaluating \sysname's true positive rates and 
false positive rates when detecting \dins. In this experiment, we fill the detected 
DMA buffers (as determined  by the \dptrs and transfer sizes) with zeros, 
instead of meaningful or fuzzer-generated data. This is to 
make our experiment precise and reproducible.

The result shows that \sysname can accurately identify \dins on
different architectures, MCU models and OSes. In other words, 
\sysname is accurate, hardware-independent and OS-agnostic. 

\point{True positive and false negative analysis}
To establish the ground truth, we looked up the DMA configuration information in the MCU data-sheets and verified it using the instrumented emulators. There were 52
firmware-executed (emulator-observed) \dstr configurations in all the firmware. We use
these as the ground truth for evaluating true positives and false negatives.  
As shown in the upper half
of Table~\ref{table:accurracy}, \sysname correctly detected 45 out of 52 \dstr
configurations. Among those detected, 33 are \dins and 12 are \douts (the latter
is irrelevant to firmware analysis and thus their buffer access/size is
omitted). \sysname  missed 7 \dstr configurations (false negatives, 4 input and
3 output DMA channels) due to the pattern limitation discussed in
\S\ref{sec:design_streamconf}. There were 22 DMA buffers accessed by the
firmware during our tests. \sysname detected 21 of them and correctly inferred
their sizes. The undetected buffer corresponds to one of the seven
configurations missed by \sysname.

It is worth noting that firmware execution 
on the emulators never accessed the
DMA buffers created by 12 \dstr configurations correctly identified by \sysname. This is due to the 
limitation of \syspp and the \mipspic. These emulators neither prioritize nor recognize DMA-related code paths in firmware. Without DMA buffers being accessed, 
DICE cannot identify \dtran sizes. We further verified that, in all the cases where
the emulators executed the corresponding DMA access code path, \sysname successfully
identified the \dtran size.

\begin{table}[]
     \centering
     \begin{tabular}{l@{}c@{}l@{}} 
     \hline
         \textbf{}  & \multirow{2}{3cm}{\textbf{Observed by emulator (ground truth)}}     & \multirow{2}{1cm}{\textbf{TP \\\sysname}}  \\ 
         &     & \\
         \hline    
         \textbf{True \dstr configurations}      &       52    &  45 (87\%)      \\
         ~~- \dins                                &       37    &  33 (89\%)      \\
         ~~- \douts                               &       15    &  12 (80\%)      \\
         Buffer accessed (size inferred)       &       22    &  21 (95\%)      \\
     \hhline{===}
        \textbf{}  & \multirow{2}{3cm}{\textbf{Observed by emulator (ground truth)}}   & \multirow{2}{1cm}{\textbf{FP \\\sysname}}  \\ 
        &     & \\
        \hline    
        \textbf{False \dstr configurations }      &    35       &  0 (0\%)      \\
        Buffer accessed  (size inferred)      &     6       &  0 (0\%)      \\             
        \hline    
     \end{tabular}
     \caption{Accuracy of DMA stream configuration detection: true positives (upper table) and false positives (lower table).
     False \dstr configurations are pointer-like 
     values written to MMIO that do not configure DMA.}
     \label{table:accurracy}
 \end{table}

\point{False positive analysis}
We instrumented the emulators to find pointer-like values written to MMIO during
firmware execution. We then used the MCU data-sheets to select those that are
unrelated to DMA configurations. We refer to them as ``false DMA stream
configurations'' and use them as the ground truth for evaluating false
positives.
As shown in the lower half of Table~\ref{table:accurracy}, 
among all the firmware tested, the emulators observed
35 pointer-like values written to the MMIO region that do not configure DMA. 
\sysname did not consider any of them as \dptrs, thus achieving a 0\% false 
positive rate.
This is because, for the heuristics to consider a value written to MMIO as a \dptr, \
the value not only needs to be in the narrow RAM range (\ie pointer-like) but also needs 
to be accompanied by another \dptr written to the adjacent MMIO location. 
As an example, 
the TIMER1 counter
register of the PIC32MZ2048EF MCU is initialized with the value 255 via an MMIO write 
operation. This value on the MIPS M-class architecture is a valid RAM
address (a pointer), which is allocated to a global variable.
\sysname observed this value, recognized it as a pointer to RAM. However, 
since no other pointer was observed to be written to the adjacent MMIO, 
 \sysname never considered this MMIO write as part of a DMA configuration. 
Among the 35 pointer-like values written to MMIO, 6 were dereferenced/accessed
by firmware. Since they were not \dptrs, \sysname did not intercept the memory
accesses or infer the buffer sizes.

Hypothetically, if firmware reads from a memory address that was falsely
identified as a DMA channel, \sysname may provide the firmware with
analyzer-provided input (\eg fuzzer-generated data). This may in turn corrupt
firmware execution. However, in our experiments, no false positive occurred thanks 
to the simple yet accurate heuristics.

\point{Runtime overhead}
The runtime overhead of \sysname is fairly low. It adds only 3.4\% on average to
the execution time of the sample firmware. The main source of overhead is the
instrumentation required to identify the \dtran size. This instrumentation
affects every LOAD instruction. A secondary source of overhead is the
instrumentation capturing the stream configurations. It only affects write
operations on the peripheral MMIO area in memory. Therefore, the overhead of
\sysname is determined by the number of LOAD instructions executed, the number
of streams configured, and the size (number of bytes) of the buffers. Table
\ref{table:overhead} illustrates the overhead on 10 firmware selected from the
entire set. These firmware were selected because they run smoothly on \syspp and
execute all the code paths related to DMA.

\begin{table}[]
\centering
\begin{tabular}{lrrr}
\hline
\textbf{Firmware} & \textbf{\syspp[s]} & \textbf{\sysname[s]} &  \textbf{Diff[\%]} \\
\hline
ADC PDC &89.0&94.1&5.7\\
ADC SW DMA &5.3&5.3&0.0\\
ADC Timer DMA &5.3&5.4&1.9\\
I2C TxRx DMA &2.8&2.8&0.0\\
SPI DMAC Slave &17.3&17.7&2.3\\
SPI Full DMA &28.2&28.9&2.5\\
SPI Full EX. DMA &13.6&15.1&11.0\\
SPI Half DMA &18.1&18.7&3.3\\
USART TxRx DMA &5.4&5.5&1.9\\
USART Full DMA &20.7&21.6&4.3\\
\hline
\end{tabular}
     \caption{Time needed for firmware execution to reach a fixed point when running on original \syspp (Col. 2) 
     and \sysname integrated on top of \syspp (Col. 3). 
     This shows the overhead that \sysname adds to \syspp.}
     \label{table:overhead}
\end{table}

\subsection{Fuzz-testing Real Firmware}
\label{sec:fuzzing}
In this experiment, we demonstrate that \sysname can effectively support dynamic
analysis on real-world firmware that uses DMA. To this end, we fuzz-tested 7
real-world firmware using \sysname integrated with \syspp. \sysname accurately
identified \dins on all 7 firmware and found 5 previously unknown bugs (none of
them were found by \syspp alone). \sysname supported fuzzing sessions for all firmware,
whereas \syspp alone failed to bootstrap a fuzzing session for 1 firmware. \sysname 
also achieved a much higher code coverage than \syspp, echoing the importance of 
DMA emulation and support during dynamic analysis.

\point{Experiment Setup}
We selected 7 real-world firmware from different sectors, ranging from
industrial IoT to consumer devices. These firmware represent diverse use cases
of DMA, such as data signal acquisition without CPU intervention and high
throughput data exchange. All these firmware contain OS/system libraries
(including scheduler, driver, interrupt service routine) and application logic.
As shown in Table~\ref{table:FirmwareBenchmark}, they are based on various MCU 
models and multiple OSes (i.e., FreeRTOS, Mbed OS,
bare-metal). We briefly describe below the firmware functionality and security
consequences of its bugs: %

\begin{table*}[t]
   \centering
    \begin{tabular}{lllll}
    \hline
    \textbf{Firmware} & \textbf{MCU} & \textbf{OS} & \textbf{Size}  & \textbf{Source} \\
    \hline
       Modbus              &STM32F303   & FreeRTOS     &  1.3MB    &   \url{https://github.com/DoHelloWorld/stm32f3_Modbus_Slave_UART-DMA-FreeRTOS}     \\
       Guitar Pedal        &STM32F303   & Mbed OS      &  2.4MB    &   \url{https://github.com/Guitarman9119/Nucleo_Guitar_Effects_Pedal}     \\
       Soldering Station   &STM32F103   & Baremetal    &  1.4MB    &   \url{https://github.com/PTDreamer/stm32_soldering_iron_controller}     \\
       Stepper Motor       &STM32F466   & Baremetal    &  1.4MB    &   \url{https://github.com/omuzychko/StepperHub}     \\
       \multirow{2}{*}{GPS Receiver}        & \multirow{2}{*}{STM32F103}  & \multirow{2}{*}{Baremetal}    &  \multirow{2}{*}{798KB}    &   \url{https://github.com/MaJerle/GPS_NMEA_parser},  \\
       &&&&\url{https://github.com/MaJerle/STM32_USART_DMA} \\
       MIDI Synthesizer    &STM32F429   & Baremetal  &  0.7MB    & \url{https://github.com/mondaugen/stm32-codec-midi-mmdsp-test} \\
       Oscilloscope        &STM32F103   & Arduino    &  0.7MB    & \url{https://github.com/pingumacpenguin/STM32-O-Scope} \\

    \hline
    \end{tabular}
    \caption{Real-world Firmware fuzz-tested with \syspp and \sysname}
    \label{table:FirmwareBenchmark}
\end{table*}

\textbf{\textit{Modbus:}} 
Modbus is a master-slave communication protocol that is widely used in
commercial PLC (Programmable Logic Controller) devices. This firmware is a
highly-optimized implementation of Modbus slave that uses DMA to receive
commands from Modbus master. As PLC devices normally control critical industrial
processes, bugs in this type of system can lead to Stuxnet-like \cite{Stuxnet}
attacks and cause physical damage. 

\textbf{\textit{Guitar Pedal:}}
This firmware includes the digital signal processing (DSP) routines for creating
musical effects in an electric guitar.
It also includes a graphical user interface (GUI) for configuring the effects.
This firmware is a typical example of mixed-signal (digital and analog)
application, which uses DMA to continuously sample analog channels at a fixed
rate. Vulnerabilities in this firmware can crash the firmware or produce
unexpected sound effects that are harmful to human hearing. 

\textbf{\textit{Soldering Station:}}
This is a customized firmware for the KSGER mini soldering station. It includes
a PID (Proportional–Integral–Derivative) temperature control routine and a
graphical user interface for configuring and operating the device. The firmware
uses DMA to read multiple ADC channels continuously in circular mode.
Vulnerabilities in this firmware can destroy the heating element and cause
injuries to operators.

\textbf{\textit{Stepper Motor:}} 
This is the firmware for the stepper motor controller in a CNC (Computer
Numerical Control) machine. CNC machines are widely used in 3D printers, drills,
lathes, etc. The firmware implements a stepper motor control routine, a command
parser, and a proprietary communication protocol. It uses DMA to achieve high
communication throughput and control speed. Vulnerabilities in these devices can
be exploited to modify the motor speed or bypass the safety checks in the CNC
machine.

\textbf{\textit{GPS Receiver:}} 
This firmware implements the GPS receiver communication protocol defined by the
National Marine Electronics Association (NMEA) \cite{NMEAgps}. This firmware
uses DMA for serial communication. We discuss this firmware as an example in
\S\ref{motivation}. Vulnerabilities in this firmware can be exploited to
manipulate navigation, and in turn, control autonomous or human-operated
vehicles.

\textbf{\textit{MIDI Synthesizer:}}
This firmware implements the MIDI protocol and controls the synthesizer operation. 
It processes stereo audio inputs through the WM8778 audio codec and digital signal 
processing routines, and outputs audio to external amplifiers. 
It uses DMA to receive and parse MIDI messages with 
low latency and overhead. Vulnerabilities in this 
firmware can crash the firmware and produce output signals that 
can overcharge the amplifiers and damage the electronic circuit of the instrument. 

\textbf{\textit{Oscilloscope:}}
This is a minimal oscilloscope that includes a touchscreen 
as the user interface and supports PC communication for data acquisition.
This firmware uses DMA for continuous and fast sampling of
electrical signals through ADC.
Vulnerabilities in this firmware can crash the system, corrupt 
the data acquired and present false information to the user.

We use unmodified AFL \cite{afl} as our fuzzer (\ie generating DMA and other
firmware input) and fuzz-tested all firmware using \sysname on \syspp. We
launched the fuzzer with random seed input and fuzz-tested each firmware for 48
hours. As pointed out by \cite{wycinwyc}, memory corruption errors are less
likely to crash the MCU firmware than computer programs, which causes fuzzer to
miss some bugs after triggering them. To mitigate this problem, we used the same
simple memory error detector (or sanitizer) described in \cite{p2im}. It grants
read+execute permission to Flash, read+write permission to RAM and the
peripheral MMIO region, and no permission to the rest of memory space. This
simple detector allows for detection of access violations that cross region
boundaries, but not those within a memory region. Besides, we implemented a more
fine-grained error detector for buffer over/under-flow detection. We inserted
red-zones before and after buffers at compile time. Accesses to red-zones will crash the
firmware execution. 
Although red-zones make bugs more visible, they are not required to launch fuzzing. 
In other words, firmware binary can be fuzz-tested ``as is" without re-compilation.
We note that an advanced memory sanitizer may help detect
more bugs in our experiment, but designing such a sanitizer is out of the scope
of this paper.

\point{Fuzzing Statistics}
For all tested firmware, \sysname was able to automatically and completely
identify  \dins (\dptr + transfer size), and feed fuzzer-generated input to DMA
buffers through RAM hooks. We did not observe any falsely identified \dins
(i.e., no false positives). On the other hand, \syspp alone was able to fuzz
test only 6 out of 7 firmware. It failed to fuzz test MIDI Synthesizer firmware 
because the firmware only consumes inputs through \dins that are
not supported by \syspp. 

As shown in Table~\ref{table:benchmark}, \sysname outperforms \syspp on 5 out of
7 firmware in terms of fuzzing coverage. \sysname improves the basic block
coverage by up to 30.4\%, and increase the number of paths triggering new execution patterns by up to 79
times. 

The improvement in path coverage is much more significant than basic block
coverage, for two reasons. First, a larger number of basic blocks in firmware
are executed during the booting process, when MCU hardware and OS are
initialized. As no DMA operation is involved during booting, these basic blocks
are reachable even without DMA support, which allows \syspp to achieve a fairly
high basic block coverage. Second, firmware code is highly reused on MCUs due to
constrained device storage. For example, the USART peripheral on the Modbus firmware is used by 
the console to print messages and by the Modbus protocol to reply to Modbus masters. 
The USART driver functions are invoked by the console, which operates without using DMA, 
and the Modbus protocol, which operates through DMA.
Supporting DMA does not significantly increase basic block coverage for these driver
functions as most of them are also used by non-DMA operations (i.e., console printing). However, without
DMA emulation or support, code paths that depend on DMA input (i.e., Modbus protocol stack) can never be
reached or tested, despite that these paths may share many basic blocks with
other paths unrelated to DMA operations or input. 

As evidenced by the result, many firmware contain a great number of DMA-related
paths (hence the drastic increase in path coverage under \sysname). Such paths
cannot be explored by analyzers without using \sysname.

Also significantly, \sysname improves the Max Depth by up to 500\% (on Stepper Motor).
This improvement indicates that, with \sysname (or generic DMA emulation),
dynamic analyzers can now dig much deeper into firmware code, unveiling states
and bugs residing at far ends of executions. This result also indicates that,
with the ability to directly feed input to DMA buffers, even off-the-shelf
fuzzers like AFL (without DMA awareness) can be used for fuzzing firmware
relying on \dins.

In terms of fuzzing speed, \sysname is slower on 3 out of 7 firmware than
\syspp, with 18\% as the worst-case slowdown (observed on Stepper Motor). The
slower fuzzing speed is not only caused by the overhead of DMA support
(discussed in \S\ref{sec:unit_test}), but also the fact that more basic blocks
and paths are executed on each fuzzer run thanks to the added DMA support. 

As 2 rare cases, fuzzing Soldering Station and Oscilloscope firmware with DMA support 
turned out to be faster, 5.6\% and 92.1\% respectively, with slightly lower code coverage. 
We found that in these cases \dins
through ADC allows rapid consumption of input data, which caused the firmware
execution to finish much earlier than without DMA support.

\begin{table*}[t]
     \scriptsize
     \centering
   
     \begin{tabular}{l|l@{ }l@{ }l|l@{ }l@{ }l|l@{ }l@{ }l|l@{ }l@{ }l|l@{ }l@{ }l|l@{ }l@{ }l|l@{ }l@{ }l}
     
     \hline

      & \multicolumn{3}{|c|}{\textbf{Modbus}} & \multicolumn{3}{|c|}{\textbf{Guitar Pedal}}  & \multicolumn{3}{|c|}{\textbf{Soldering St.}} & \multicolumn{3}{|c|}{\textbf{Stepper Motor}} & \multicolumn{3}{|c}{\textbf{GPS Receiver}}  & \multicolumn{3}{|c}{\textbf{MIDI Synth.}}  & \multicolumn{3}{|c}{\textbf{Oscilloscope}} \\ 
     \hline
     \textbf{Framework}  & \syspp & \sysname & $\Delta$\%  & \syspp & \sysname & $\Delta$\% & \syspp & \sysname & $\Delta$\% & \syspp & \sysname & $\Delta$\% & \syspp & \sysname & $\Delta$\% & \syspp & \sysname & $\Delta$\% & \syspp & \sysname & $\Delta$\% \\
     
     \hline
     \textbf{BBL Cov. [\%]} & 
52.6 & 58.7 & 11.6 & 
16.9 & 17.0 & 0.6 & 
31 & 31 & 0 & 
22.3 & 25.6 & 14.8 & 
11.5 & 15.0 & 30.4 &
0.0 & 40.8 & N/A &
27.3 & 27.3 & 0.0 \\

     \textbf{Total Paths} & 
16 & 1276 & 7875 & 
3267 & 3773 & 15.5 & 
172 & 166 & -3.5 & 
4595 & 5276 & 14.8 & 
30 & 1988 & 6527 &
0 &  588  & N/A &
618 &  590  & -4.5 \\ 

     \textbf{Max Depth} & 
2 & 8 & 300 & 
4 & 5 & 25 & 
3 & 3 & 0 & 
2 & 12 & 500 & 
5 & 6 & 20 &
0 & 3 & N/A &
5 & 4 & -20 \\

     \textbf{Speed [run/s]} & 
41.6 & 41.0 & -1.4 & 
3.8 & 3.8 & 0 & 
17.9 & 18.9 & 5.6 & 
22.2 & 18.2 & -18.0 &   
49.4 & 48.9 & -1.0  &
0 & 59.9 & N/A  &
0.76 & 1.46 & 92.1  
\\
     \hline
          
     \end{tabular}
     \caption{Statistics of fuzz-testing real firmware using \sysname on \syspp. 
     \syspp was unable to fuzz test MIDI Synthesizer firmware which solely consumes input through \dins.}
     \label{table:benchmark} 
\end{table*}

\point{Detected New Bugs and Case Study}
\sysname found 5 unique, previously unknown bugs that
\syspp alone (i.e. without DMA emulation) cannot detect (Table~\ref{table:bugs}).
We manually examined these bugs (3 in Modbus and 2 in MIDI Synthesizer) and 
confirmed that it is the DMA support that makes \sysname outperforms existing 
dynamic analysis frameworks, such as \syspp. More specifically, Modbus and the MIDI
Synthesizer firmware receive commands from the USART peripheral
through DMA. Without DMA support, no command can be received by the firmware. As
a result, the command parsing logic and application logic, where all 5 bugs were found, 
can never be executed.

\begin{table}
  \centering
   \begin{tabular}{m{1cm} m{0.9cm} m{1.9cm} m{3.3cm}} 
   \hline
   \textbf{Firmware} & \textbf{Bug ID} & \textbf{Bug type} & \textbf{Security consequences} \\ 
   \hline
   \multirow{3}{1cm}{Modbus} &\multirow{2}{1cm}{1, 2}  & \multirow{2}{1.9cm}{Buffer overwrite}   &\multirow{2}{3.3cm}{Corrupt data structure with\\attacker controlled values.}\\
      &&&  \\
      & 3                     & Buffer overread              &Information leakage. \\      
   \hline
   \multirow{2}{1cm}{MIDI \\Synthesizer} & \multirow{2}{1cm}{4, 5} & \multirow{2}{1.9cm}{Free of memory not on the Heap} & \multirow{2}{3.3cm}{Firmware crash,\\denial-of-service.} \\
   &&& \\
   \hline
   \end{tabular}
   \caption{New bugs found by \sysname in Modbus and MIDI Synthesizer firmware. None of them can be found by \syspp alone.}
   \label{table:bugs}
\end{table}

We verified that all the bugs are real and reproducible on real devices with
the same fuzzer-generated input. All the bugs detailed in Table \ref{table:bugs}
are remotely exploitable. They are triggered by the commands that the 
firmware receives through the USART peripheral via a \din.

The bugs found in the MIDI Synthesizer firmware (ID 4 and 5) are caused by freeing 
memory not on the heap. The firmware stores the MIDI messages received through \dins in 
the buffers that are either statically allocated (as global variables) or 
dynamically allocated on the heap. When the buffers are allocated on the heap, the 
firmware invokes free() function to deallocate them after the MIDI messages is 
processed. In these bugs, the firmware uses global buffers, but still 
invokes the free() function which is supposed to free only buffers allocated on 
heap. This causes that memory not on the heap is freed, which 
may lead to firmware crash and denial-of-service. These bugs cannot be found by 
\syspp alone because the free() function is only invoked when the MIDI 
messages received from \dins are processed.

For the Modbus firmware, \sysname identified 2 buffer overwrite and 
1 buffer overread bugs. The root cause 
is improper validations of array indexes. Specifically,
the firmware uses untrusted input for calculating array indexes, but fails to
validate the computed indexes or ensure the indexes are referencing valid positions
within the arrays. Modbus receives commands from a shared fieldbus in a typical 
PLC device setup. Therefore, any malicious/compromised device
connected to the fieldbus can exploit these bugs by sending crafted commands. These
bugs allow an attacker to corrupt data structures or retrieve secrets stored in
the firmware memory, e.g. critical parameters of the PLC control routine.

Listing \ref{listing:1} shows the code snippet of Bug \#1. \texttt{startAddr} is
calculated from \texttt{modbusRxTxBuffer[]}, which is the DMA buffer that holds
untrusted input (Line 265). The firmware checks if \texttt{startAddr} is within
the valid range of array \texttt{modbusMemory[]} (Line 266), and then uses it as
an index for array access (Line 270). The input validation at
Line 266 is wrong (the correct check should be \texttt{startAddr >=
MODBUS\_SLAVE\_REGISTERS\_NUM}), which causes buffer overwrite at Line 270.

\lstset{style=mystyle}
\begin{minipage}{\linewidth}
\begin{lstlisting}[language=C,caption=Code snippet of Bug \#1, firstnumber=265, label=listing:1]
uint16_t startAddr = modbusRxTxBuffer[2] << 8 | modbusRxTxBuffer[3];
if(startAddr > MODBUS_SLAVE_REGISTERS_NUM) // improper input validation
  answerLen = modbusSlaveErrorSet(0x02);
else
{
  modbusMemory[startAddr] = modbusRxTxBuffer[4] << 8 | modbusRxTxBuffer[5]; // buffer overwrite
  answerLen = modbusRxCount;
}
\end{lstlisting}
\end{minipage}

We also investigated the potential reasons for our experiment not finding bugs
in the other five firmware. In general, fuzz-testing firmware with \sysname on
\syspp faces the open challenges as with other existing tools, such as the lack
of error detectors/sanitizers for MCU, limited ability to solve complex state
machines and path constraints, etc., which are out of the scope for this paper.
We also identified the following reasons specific to DMA.

First, some firmware using DMA can quickly drain fuzzer input, without going
deep into the code. Such firmware tends to allocate large DMA buffers for
high-throughput data transfers. For example, Stepper Motor uses 2 buffers of 4
KB for receiving and transmitting data. However, AFL prefers to generate short
input sequences to achieve better fuzzing performance. 

Second, some \dins do not directly influence firmware control flow, but \sysname
still treats them as ``risky'' channels and lets the fuzzer generate and mutate
inputs for them. For example, Soldering Station, Guitar Pedal and Oscilloscope 
continuously sample analog input using DMA and performs mathematical calculations on the
sampled inputs. These DMA inputs can rarely change firmware execution paths or
trigger bugs. However, due to the design requirement of being
peripheral-agnostic, \sysname cannot detect or exclude such fuzzing-unworthy
\dins, and thus, ends up spending too much time on them, instead of focusing on other
fuzzing-worthy \dins.

\section{Discussion}
\label{sec:discussion}

\subsection{Location of \ddesr}
\sysname identifies the \dins based on the assumption that \ddesrs are always written to \dctr through MMIO operations. However, in some rare cases, \ddesrs are stored in RAM, which are not supported by \sysname. 
We admit it a limitation of \sysname. 
To measure how prevalent the \ddesrs are stored in RAM, we surveyed the complete STM32 MCU portfolio of ST Microelectronics, a top-five MCU vendor according to \cite{TheMacleanReport2017} and the most popular MCU vendor in terms of the number of Github repositories (19,870 unique entries by Nov. 2019). 
The product lines we analyzed include ultra-low-power, main-stream, and high-performance families. 
Our analysis showed that all MCUs (983 in total) store \ddesrs in the peripheral memory area, 
while the STM32H7 product line (59 MCUs) also allows storing \ddesrs in RAM.
To conclude, only 6\% of STM32 MCUs can optionally store \ddesrs in RAM, and therefore, it is acceptable for \sysname not to handle this rare case.

\subsection{DMA Buffer Size Identification}
Unlike source and destination, buffer size cannot be reliably identified from a \dstr configuration event. 
To solve this, \sysname adopts a conservative heuristic for buffer size identification. 
The heuristic, which gradually expands the perceived DMA buffers at memory read that falls right after the buffer boundary, may produce a smaller-than-actual size upon memory reads that are not consecutive in space. 
This inaccuracy, although possible in theory, is not observed in our evaluation. 
Moreover, when the firmware executes for long enough, \sysname may progressively identify the correct buffer size.
Therefore, it is reasonable to use such a conservative heuristic which trades identification accuracy for the firmware stability (when a wrongly identified DMA buffer byte is read, the firmware can crash).

\subsection{Architecture beyond ARM and MIPS}
\label{sec:more_arch}
\sysname can be applied to other architectures that meet three
requirements: 
(R1) the architecture uses designated memory regions for peripherals (MMIO), Flash and RAM; 
(R2) \dstrs configurations are written to the peripheral region via MMIO;
(R3) \dtrans follow the life cycle depicted in Figure \ref{fig:DMALifeCycle}.

We analyzed RISC-V, the increasingly popular architecture used in MCUs. 
We confirmed that RISC-V meets these requirements, and therefore, is compatible
with \sysname. Specifically, we studied the data-sheets of the RISC-V GD32VF103 MCU
 \cite{GD32VF103RISCV32}. The MCU uses separate memory regions for 
peripherals, RAM and Flash (0x40000000-0x5003FFFF, 0x20000000-0x20017FFF, and 
0x08000000-0x0801FFFF respectively), thus meeting R1. 
It also writes \dstrs configurations via MMIO to peripherals and meet R2. 
The MCU obeys the DMA life cycle illustrated in Figure \ref{fig:DMALifeCycle} 
and meet R3. 
This result demonstrates that \sysname
is generically applicable to at least three different architectures, namely ARM, MIPS, 
and RISC-V.

\subsection{Devices beyond MCUs}
\label{sec:dev_b_mcu}
\sysname is designed to support DMA in firmware analysis for MCU devices. It
solves multiple challenges, especially, the hardware and software diversity of
MCUs. Other platforms such as desktop and mobile devices use similar DMA \ddesrs
and follow the same DMA life cycle as MCUs. However, those more powerful
platforms frequently store \ddesrs in RAM, rather than in \dctr registers
(MMIO), for flexibility reasons (e.g., to support complex concatenated \dtrans).
As we discussed in \S \ref{sec:design}, RAM-stored \ddesrs are not supported by
\sysname. Therefore,  \sysname cannot be directly used for emulating DMA on
platforms other than MCUs.   

Existing work such as PeriScope~\cite{periscope} can handle DMA on
Linux-based platforms during dynamic analysis. PeriScope instruments Linux
kernel DMA APIs to monitor the creation and destroy of \dins and to manipulate
DMA input. PeriScope is OS-specific. Unlike \sysname,
PeriScope is not applicable to MCUs, which have highly diverse OS and firmware, and therefore, pose unique challenges for DMA emulation.

\subsection{Integration with Other Firmware Analyzers}

The design of \sysname allows for easy integration with various firmware analyzers, 
providing them with the capability of analyzing firmware that use DMA.
As a demonstration, we integrated \sysname with Avatar2 \cite{avatar2}, a flexible dynamic firmware analysis framework. The implementation is only 240 lines of Python code. 
This integration allows Avatar2 to recognize and manipulate DMA data when it was
read by the firmware, and in turn, to analyze firmware that uses DMA for input.

Furthermore, we integrate \sysname with Symbion \cite{gritti2020symbion} (an extension to the Angr 
\cite{angrShoshitaishvili2016} framework). This integration allows DMA-aware concolic execution on MCU firmware. 
Specifically, we used \sysname and Avatar2 to identify the DMA buffers used by a 
firmware as the firmware runs on a real development board. Upon reaching an
interesting point of analysis, Symbion moves the concrete state of this execution to Angr. Thanks to \sysname, Angr can now recognize and symbolize the DMA buffers in the concrete
state, achieving a more precise and comprehensive concolic execution.

\subsection{Peripherals beyond DMA controller}
Many merits of \sysname, such as hardware-independent and firmware compatibility, are attributed to our abstraction of the \din. 
We believe that for other peripherals (e.g., counters and comparators), which have well-defined functionalities and identifiable configurations like \dctrs do, can be abstracted in a similar way.
We admit that identifying if and how a peripheral can be abstracted for dynamic analysis purposes, in a hardware-independent and firmware-compatible way, is an interesting research topic that we would like to pursue in the future.

\section{Related Work}
\subsection{Dynamic Firmware Analysis}
Multiple existing works tackled the challenging problem of dynamic firmware analysis. 
They are divided into hardware-in-the-loop emulation approaches and full emulation approaches by whether real devices are required in the process of dynamic analysis. 
Avatar \cite{avatar} proposed a novel hardware-in-the-loop emulation mechanism, which forwards peripheral operations to a real device while executing the firmware in the emulator. 
It conducted concolic execution for MCU firmware. 
Surrogates \cite{surrogates} improved the performance of peripheral operation forwarding by customized hardware.
\cite{wycinwyc} fuzz-tested simple programs with artificially-implanted bugs using Avatar, which demonstrated that memory corruption vulnerabilities are much less likely to crash on MCU than on desktop. 
Avatar2 \cite{avatar2} extended Avatar with the record and replay capability for the forwarded peripheral operations. 
Charm \cite{charm} fuzz-tested Android device drivers by a hardware-in-the-loop emulator that uses a similar forwarding technique with Avatar. 
Prospect \cite{prospect} forwarded peripheral operations made through syscalls, the abstraction of which is not available for MCU devices. 
\cite{peri_caching} combined runtime program state approximation with peripheral access caching to facilitate dynamic analysis.  

Hardware-in-the-loop emulation approaches suffer from poor performance and scalability due to the slow forwarding speed and one-to-one binding between emulator instances and real devices.
Several recent works addressed this by removing the need for real devices through full emulation. 
Pretender \cite{pretender} generated approximated peripheral models from the peripheral operations that are forwarded to the real device by Avatar \cite{avatar}.
With the model, it successfully executed and fuzzed-tested several simple firmware with manually-injected vulnerabilities without using any real device. 
P\textsuperscript{2}IM \cite{p2im} completely removed the usage of real devices by automatically modeling the processor-peripheral interfaces while emulating the firmware. 
It fuzz-tested several real-world firmware of typical embedded applications and found real bugs. 
HALucinator~\cite{HALucinator} adopts a high-level emulation-based approach 
which replaces Hardware Abstraction Layer (HAL) functions with manually-crafted 
handlers by library matching on binary. It fuzz-tested network stack, file 
system, serial port and PLC, and found real bugs. 
PartEmu \cite{partemu} fuzz-tested ARM TrustZone software stack by a hybrid approach of 
replacing software components with stubs and modeling peripheral hardware with manually crafted register value patterns. 
Various works dynamically analyzed Linux-based firmware by full emulation \cite{firmadyne,dyn_webif,symdrive}. 
Those firmware are more similar to general-purpose desktop software than truly embedded firmware. 
Emulators have much better support for Linux-based firmware, which uses less diverse peripherals than MCU firmware. 
However, none of these works, either hardware-in-the-loop emulation or full emulation, were able to dynamically analyze MCU firmware using DMA without requiring any source code.
The major obstacle is the un-emulated DMA controllers, which are vendor-specific and possibly proprietary. 
Our work can extend both hardware-in-the-loop (Avatar) and full emulation (P\textsuperscript{2}IM \cite{p2im}) mechanism with DMA support and conduct dynamic analysis. 
Although HALucinator~\cite{HALucinator} can analyze DMA-enabled firmware (because its high-level 
emulation totally removes DMA operations), their approach (specifically, the 
library matching component) requires source code of HAL. Neither can HALucinator
find bugs in firmware components that are replaced by the high-level emulation, 
such as drivers for \dctr and other peripherals. \sysname adopts a completely 
different approach which identifies DMA input channels from unmodified 
firmware. Therefore, \sysname is able to find bugs in the whole firmware stack, 
without requiring any source code or manually created handlers.

\subsection{DMA Attacks \& Analyses}
Various works revealed attacks enabled by DMA. 
To name a few, PCILeech \cite{PCILeech} revealed that malicious peripherals with DMA capability (e.g., PCIe peripherals) can access/modify arbitrary physical memory addresses and gain full control over the victim computer if IOMMU is not enabled (IOMMU enables virtual memory for I/O devices).
Thunderclap \cite{thunderclap} further demonstrated that DMA attacks are still feasible even with IOMMU enforced. 
These attacks, however, are not feasible on MCUs because MCU peripherals normally do not have DMA capability, and DMA is conducted through a dedicated DMA controller which is part of the System-on-Chip (SoC) and considered trusted. 
PeriScope \cite{periscope} identified a compromised peripheral device (e.g., Wi-Fi chip) can attack kernel device drivers by sending malicious input through DMA, and therefore, fuzz-tested the DMA channels. 
It identifies DMA channels by instrumenting Linux kernel APIs, which however is not applicable to bare-metal MCU firmware. 

\section{Conclusion}

We presented a survey showing the prevalence and diverse usages of DMA on
MCU-based embedded devices. We highlighted the importance of supporting
DMA-enabled peripherals during dynamic firmware analysis. To address existing
firmware analyzers' inability to test DMA-enabled firmware, we designed and
built \sysname, a drop-in solution that enables analyzer-generic and
hardware-independent emulation of \dins. By identifying and observing DMA
configurations and accesses by firmware during emulated execution, \sysname
detects \dins that are dynamically created by firmware. It also dynamically
infers the locations and sizes of memory buffers used as DMA transfer
destinations. Without requiring any human assistance or firmware source code,
\sysname allows firmware analyzers to run and test DMA-related code, and in
turn, find bugs or vulnerabilities in firmware that otherwise cannot be reached
or triggered.

We integrated \sysname into \syspp (for ARM Cortex-M) and a MIPS \mipspic. We evaluated \sysname using 
83 sample firmware and 7 real-world firmware. Its runtime overhead is 
low (3.4\%) and its emulation accuracy is very
high (89\% true positive rate and 0\% false positive rate). When used for fuzzing the real-world firmware, \sysname increased
code path coverage by as much as 79X. Moreover, it helps detect 5 unique,
previously unreported bugs, which would not have been found without the generic and automatic DMA emulation.  

\section*{Acknowledgment}
The authors would like to thank the anonymous reviewers for their insightful
comments. This project was supported by 
the National Science Foundation (Grant\#: CNS-1748334), %
the Office of Naval Research (Grant\#: N00014-18-1-2043), %
and the Army Research Office (Grant\#: W911NF-18-1-0093). %
Any opinions, findings, and conclusions or recommendations expressed in this
paper are those of the authors and do not necessarily reflect the views of the
funding agencies.

\bibliographystyle{plain}
\bibliography{refs}

\begin{thebibliography}{10}

\bibitem{nRF52832datasheet}
Nrf52832 datasheet.
\newblock \url{https://infocenter.nordicsemi.com/pdf/nRF52832_PS_v1.4.pdf}.
\newblock Accessed: Sep 2019.

\bibitem{HALucinator}
Halucinator: Firmware re-hosting through abstraction layer emulation.
\newblock In {\em 29th {USENIX} Security Symposium}, 2020.

\bibitem{dec_pdp-8}
C.~Gordon Bell, Allen Newell, and Daniel~P. Siewiorek.
\newblock Structural levels of the pdp-8.
\newblock
  \url{http://digitalcollections.library.cmu.edu/awweb/awarchive?type=file&item=356992},
  1978.
\newblock Accessed: Apr 2020.

\bibitem{qemu_sp}
Fabrice Bellar.
\newblock Qemu features/softmmu.
\newblock \url{https://wiki.qemu.org/Features/SoftMMU}.
\newblock Accessed: Apr 2020.

\bibitem{AutomotiveCheckoway}
Stephen Checkoway, Damon McCoy, Brian Kantor, Danny Anderson, Hovav Shacham,
  Stefan Savage, Karl Koscher, Alexei Czeskis, Franziska Roesner, and Tadayoshi
  Kohno.
\newblock Comprehensive experimental analyses of automotive attack surfaces.
\newblock In {\em 20th {USENIX} Security Symposium, San Francisco, CA, USA,
  August 8-12, 2011, Proceedings}. {USENIX} Association, 2011.

\bibitem{firmadyne}
Daming~D Chen, Maverick Woo, David Brumley, and Manuel Egele.
\newblock Towards automated dynamic analysis for linux-based embedded firmware.
\newblock In {\em Network and Distributed System Security Symposium (NDSS)},
  2016.

\bibitem{inception}
Nassim Corteggiani, Giovanni Camurati, and Aur{\'e}lien Francillon.
\newblock Inception: system-wide security testing of real-world embedded
  systems software.
\newblock In {\em 27th {USENIX} Security Symposium}, 2018.

\bibitem{dyn_webif}
Andrei Costin, Apostolis Zarras, and Aur{\'e}lien Francillon.
\newblock Automated dynamic firmware analysis at scale: a case study on
  embedded web interfaces.
\newblock In {\em ACM Asia Conference on Computer and Communications Security},
  2016.

\bibitem{chibiOS}
Geovanny Di~Sirio.
\newblock {ChibiOS}.
\newblock \url{http://chibios.org}, 2017.
\newblock Accessed: Sep 2019.

\bibitem{p2im}
Bo~Feng, Alejandro Mera, and Long Lu.
\newblock P2im: Scalable and hardware-independent firmware testing via
  automatic peripheral interface modeling.
\newblock In {\em 29th {USENIX} Security Symposium}, 2020.

\bibitem{PCILeech}
U.~Frisk.
\newblock Direct memory attack the kernel.
\newblock In {\em Proceedings of DEFCON’24}, 2016.

\bibitem{Gartner2020}
Gartner.
\newblock Gartner says 5.8 billion enterprise and automotive iot endpoints will
  be in use in 2020.
\newblock
  \url{https://www.gartner.com/en/newsroom/press-releases/2019-08-29-gartner-says-5-8-billion-enterprise-and-automotive-io}.
\newblock Accessed: Nov 2019.

\bibitem{NMEAgps}
GPSinformation.org.
\newblock {NMEA} data.
\newblock \url{https://www.gpsinformation.org/dale/nmea.htm}.
\newblock Accessed: Sep 2019.

\bibitem{gritti2020symbion}
Fabio Gritti, Lorenzo Fontana, Eric Gustafson, Fabio Pagani, Andrea Continella,
  Christopher Kruegel, and Giovanni Vigna.
\newblock Symbion: Interleaving symbolic with concrete execution.
\newblock In {\em Proceedings of the IEEE Conference on Communications and
  Network Security (CNS)}, June 2020.

\bibitem{pretender}
Eric Gustafson, Marius Muench, Chad Spensky, Nilo Redini, Aravind Machiry,
  Yanick Fratantonio, Davide Balzarotti, Aurelien Francillon, Yung~Ryn Choe,
  Christophe Kruegel, et~al.
\newblock Toward the analysis of embedded firmware through automated
  re-hosting.
\newblock In {\em International Symposium on Research in Attacks, Intrusions
  and Defenses (RAID) 2019)}, 2019.

\bibitem{HardinSPHSK18}
Taylor Hardin, Ryan Scott, Patrick Proctor, Josiah~D. Hester, Jacob Sorber, and
  David Kotz.
\newblock Application memory isolation on ultra-low-power mcus.
\newblock In {\em 2018 {USENIX} Annual Technical Conference, {USENIX} {ATC}
  2018, Boston, MA, USA, July 11-13, 2018}, pages 127--132, 2018.

\bibitem{partemu}
Lee Harrison, Hayawardh Vijayakumar, Rohan Padhye, Koushik Sen, and Michael
  Grace.
\newblock {PARTEMU}: Enabling dynamic analysis of real-world trustzone software
  using emulation.
\newblock In {\em 29th {USENIX} Security Symposium}, 2020.

\bibitem{triforceafl}
Jesse Hertz and Tim Newsham.
\newblock Triforceafl.
\newblock
  \url{https://www.nccgroup.trust/us/about-us/newsroom-and-events/blog/2016/june/project-triforce-run-afl-on-everything/}.
\newblock Accessed: Sep 2019.

\bibitem{TheMacleanReport2017}
IC~Insights.
\newblock The mcclean report 2017 - april update.
\newblock
  \url{https://www.eenewsanalog.com/news/ma-moves-alter-mcu-vendor-ranking-0 }.
\newblock Accessed: Sep 2019.

\bibitem{peri_caching}
Markus Kammerstetter, Daniel Burian, and Wolfgang Kastner.
\newblock Embedded security testing with peripheral device caching and runtime
  program state approximation.
\newblock In {\em 10th International Conference on Emerging Security
  Information, Systems and Technologies (SECUWARE)}, 2016.

\bibitem{prospect}
Markus Kammerstetter, Christian Platzer, and Wolfgang Kastner.
\newblock Prospect: peripheral proxying supported embedded code testing.
\newblock In {\em ACM Symposium on Information, Computer and Communications
  Security}, 2014.

\bibitem{surrogates}
Karl Koscher, Tadayoshi Kohno, and David Molnar.
\newblock Surrogates: Enabling near-real-time dynamic analyses of embedded
  systems.
\newblock In {\em WOOT}, 2015.

\bibitem{thunderclap}
A~Theodore Markettos, Colin Rothwell, Brett~F Gutstein, Allison Pearce, Peter~G
  Neumann, Simon~W Moore, and Robert~NM Watson.
\newblock Thunderclap: Exploring vulnerabilities in operating system iommu
  protection via dma from untrustworthy peripherals.
\newblock In {\em Network and Distributed System Security Symposium (NDSS)},
  2019.

\bibitem{MICROCHIPproductportfolio}
Microchip.
\newblock Microchip product portfolio march 2019.
\newblock
  \url{https://www.microchip.com/ParamChartSearch/chart.aspx?branchID=30063},
  2019.
\newblock Accessed: March 2019.

\bibitem{MICROCHIP16BITS}
Microchip.
\newblock Quick reference guide 16bit microcontrollers.
\newblock \url{http://ww1.microchip.com/downloads/en/DeviceDoc/30010109F.pdf},
  2019.
\newblock Accessed: March 2019.

\bibitem{MICROCHIP32BITS}
Microchip.
\newblock Quick reference guide 32bit microcontrollers.
\newblock \url{http://ww1.microchip.com/downloads/en/DeviceDoc/60001455D.pdf},
  2019.
\newblock Accessed: March 2019.

\bibitem{MICROCHIP8BITS}
Microchip.
\newblock Quick reference guide 8bit microcontrollers.
\newblock \url{http://ww1.microchip.com/downloads/en/DeviceDoc/30009630M.pdf},
  2019.
\newblock Accessed: March 2019.

\bibitem{RemoteCarHackMiller}
Valasek~Chris Miller~Charlie.
\newblock Remote exploitation of an unaltered passenger vehicle.
\newblock \url{http://illmatics.com/Remote Car Hacking.pdf}.
\newblock Accessed: Nov 2019.

\bibitem{avatar2}
Marius Muench, Dario Nisi, Aur{\'e}lien Francillon, and Davide Balzarotti.
\newblock Avatar 2: A multi-target orchestration platform.
\newblock In {\em BAR}, 2018.

\bibitem{wycinwyc}
Marius Muench, Jan Stijohann, Frank Kargl, Aur{\'e}lien Francillon, and Davide
  Balzarotti.
\newblock What you corrupt is not what you crash: Challenges in fuzzing
  embedded devices.
\newblock In {\em Network and Distributed System Security Symposium (NDSS)},
  2018.

\bibitem{WifiSelianiMarvellAvastar}
{NIST}.
\newblock {CVE-2019-6496}.
\newblock \url{https://nvd.nist.gov/vuln/detail/CVE-2019-6496}, 2019.
\newblock Accessed: April 2020.

\bibitem{ToyotaKillerFirmware}
Koopman Phil.
\newblock A case study of toyota unintended acceleration and software safety.
\newblock
  \url{https://users.ece.cmu.edu/~koopman/pubs/koopman14_toyota_ua_slides.pdf},
  2014.
\newblock Accessed: November 2019.

\bibitem{WifiPzero}
Google Project~Zero.
\newblock Over the air: Exploiting broadcom’s wi-fi stack.
\newblock
  \url{https://googleprojectzero.blogspot.com/2017/04/over-air-exploiting-broadcoms-wi-fi_4.html},
  2017.
\newblock Accessed: November 2019.

\bibitem{symdrive}
Matthew~J Renzelmann, Asim Kadav, and Michael~M Swift.
\newblock Symdrive: Testing drivers without devices.
\newblock In {\em OSDI}, 2012.

\bibitem{GD32VF103RISCV32}
GigaDevice Semiconductor.
\newblock Gd32vf103 user manual.
\newblock
  \url{http://gd32mcu.21ic.com/data/documents/shujushouce/GD32VF103_User_Manual_EN_V1.2.pdf},
  2019.
\newblock Accessed: Nov 2019.

\bibitem{PIC32emulator}
Vakulenko Serge.
\newblock Qemu for mips pic32.
\newblock \url{https://github.com/sergev/qemu/wiki}.
\newblock Accessed: April 2020.

\bibitem{SWATT}
A.~{Seshadri}, A.~{Perrig}, L.~{van Doorn}, and P.~{Khosla}.
\newblock Swatt: software-based attestation for embedded devices.
\newblock In {\em IEEE Symposium on Security and Privacy, 2004. Proceedings.
  2004}, pages 272--282, May 2004.

\bibitem{angrShoshitaishvili2016}
Yan Shoshitaishvili, Ruoyu Wang, Christopher Salls, Nick Stephens, Mario
  Polino, Audrey Dutcher, John Grosen, Siji Feng, Christophe Hauser,
  Christopher Kruegel, and Giovanni Vigna.
\newblock {SoK: (State of) The Art of War: Offensive Techniques in Binary
  Analysis}.
\newblock In {\em IEEE Symposium on Security and Privacy}, 2016.

\bibitem{periscope}
Dokyung Song, Felicitas Hetzelt, Dipanjan Das, Chad Spensky, Yeoul Na, Stijn
  Volckaert, Giovanni Vigna, Christopher Kruegel, Jean-Pierre Seifert, and
  Michael Franz.
\newblock Periscope: An effective probing and fuzzing framework for the
  hardware-os boundary.
\newblock In {\em Network and Distributed System Security Symposium (NDSS)},
  2019.

\bibitem{STM32F4}
STmicroelectronics.
\newblock Stm32f4 reference manual.
\newblock \url{https://www.st.com/resource/en/reference_manual/dm00031020.pdf},
  2019.
\newblock Accessed: Nov 2019.

\bibitem{sun2018oat}
Zhichuang Sun, Bo~Feng, Long Lu, and Somesh Jha.
\newblock Oat: Attesting operation integrity of embedded devices.
\newblock In {\em 2020 IEEE Symposium on Security and Privacy (SP)}. IEEE,
  2020.

\bibitem{charm}
Seyed Mohammadjavad~Seyed Talebi, Hamid Tavakoli, Hang Zhang, Zheng Zhang,
  Ardalan~Amiri Sani, and Zhiyun Qian.
\newblock Charm: Facilitating dynamic analysis of device drivers of mobile
  systems.
\newblock In {\em 27th {USENIX} Security Symposium}, 2018.

\bibitem{Stuxnet}
Wikipedia.
\newblock Stuxnet.
\newblock \url{https://en.wikipedia.org/wiki/Stuxnet}, 2010.
\newblock Accessed: Sep 2019.

\bibitem{avatar}
Jonas Zaddach, Luca Bruno, Aurelien Francillon, and Davide Balzarotti.
\newblock Avatar: A framework to support dynamic security analysis of embedded
  systems' firmwares.
\newblock In {\em Network and Distributed System Security Symposium (NDSS)},
  2014.

\bibitem{afl}
Michal Zalewski.
\newblock american fuzzy lop.
\newblock \url{http://lcamtuf.coredump.cx/afl/}.
\newblock Accessed: Sep 2019.

\end{thebibliography}

\end{document}